\documentclass[aps,pre,superscriptaddress,showpacs,floats,twocolumn,amssymb,floatfix]{revtex4-1}

\input epsf
\usepackage{bm}
\usepackage{amsmath,amsfonts}
\usepackage{graphicx}
\usepackage{color}
\usepackage[usenames,dvipsnames,svgnames,table]{xcolor}
\usepackage{hyperref}
\usepackage{setspace}
\usepackage{soul}
\usepackage{accents}
\usepackage{mathtools}

\newcommand{\beq}{\begin{equation}}
\newcommand{\eeq}{\end{equation}}
\newcommand{\beqa}{\begin{eqnarray}}
\newcommand{\eeqa}{\end{eqnarray}}
\newcommand{\beqan}{\begin{eqnarray*}}
\newcommand{\eenan}{\end{eqnarray*}}

\newcommand{\ubar}{}

\newcommand{\CU}{{ U}}

\newcommand{\CR}{{\cal R}}

\newcommand{\sw}{{\xi}}
\newcommand{\swb}{{\bm \sw}}

\newcommand{\Runcorr}{\tilde{\CR}^{\scriptstyle\text{uncr}}}
\newcommand{\Rcorr}{\tilde{\CR}^{\scriptstyle\text{corr}}}

\newcommand{\jz}{{\cal J}_z}

\newcommand{\ppz}{{\cal P}_z^\perp}

\newcommand{\km}{ {\cal K}}

\newcommand{\unitx}{\hat{x}}
\newcommand{\unity}{\hat{y}}

\newcommand{\Dort}{ D_{\scriptstyle\text{o}} }

\newcommand{\Deff}{ D_{\scriptstyle\text{eff}} }

\newcommand{\eqdef}{ {\kern 0.2em}\equiv{\kern -0.5em}:{\kern 0.2 em} }
\newcommand{\defeq}{ {\kern 0.2em}:{\kern -0.5em}\equiv{\kern 0.2 em} }

\newcommand{\tauxi}{\tau_{_\xi}}
\newcommand{\tauv}{\tau_v}
\newcommand{\tautheta}{\tau_{\scriptscriptstyle{\theta}}}
\newcommand{\varv}{\mathrm{var}(v_s)}

\newcommand{\Const}{{\scriptstyle\text{mean}}}
\newcommand{\Fluct}{{\scriptstyle\text{flct}}}
\newcommand{\Dcal}{\Phi}

\begin{document}

\addtolength{\topmargin}{10pt}

\title{
Gaussian Memory in Kinematic Matrix Theory for Self-Propellers}

\author{Amir Nourhani} 
\email{nourhani@psu.edu}
\affiliation{Department of Physics, The Pennsylvania State University, University Park, PA 16802}
\author{Vincent H. Crespi}
\affiliation{Department of Physics, The Pennsylvania State University, University Park, PA 16802}
\affiliation{Department of Materials Science and Engineering, The Pennsylvania State University, University Park, PA 16802}
\affiliation{Department of Chemistry, The Pennsylvania State University, University Park, PA 16802}
\author{Paul E. Lammert}
\affiliation{Department of Physics, The Pennsylvania State University, University Park, PA 16802} 


\begin{abstract}
We extend the kinematic matrix (``kinematrix'') formalism [Phys. Rev. E {\bf 89}, 062304 (2014)], 
which via simple matrix algebra accesses ensemble properties of self-propellers
influenced by uncorrelated noise, 
to treat Gaussian correlated noises. 
This extension brings into reach many real-world biological and
biomimetic self-propellers for which inertia is significant.
Applying the formalism, we analyze in detail ensemble behaviors of a 2D 
self-propeller with velocity fluctuations and orientation evolution driven by
an Ornstein-Uhlenbeck process. On the basis of exact results, a variety of
dynamical regimes determined by the inertial, speed-fluctuation, orientational
diffusion, and emergent disorientation time scales are delineated and discussed.
\end{abstract}
\maketitle 

\section{Introduction}

\begin{figure}[b]
\begin{center}
\includegraphics[width=3.2in]{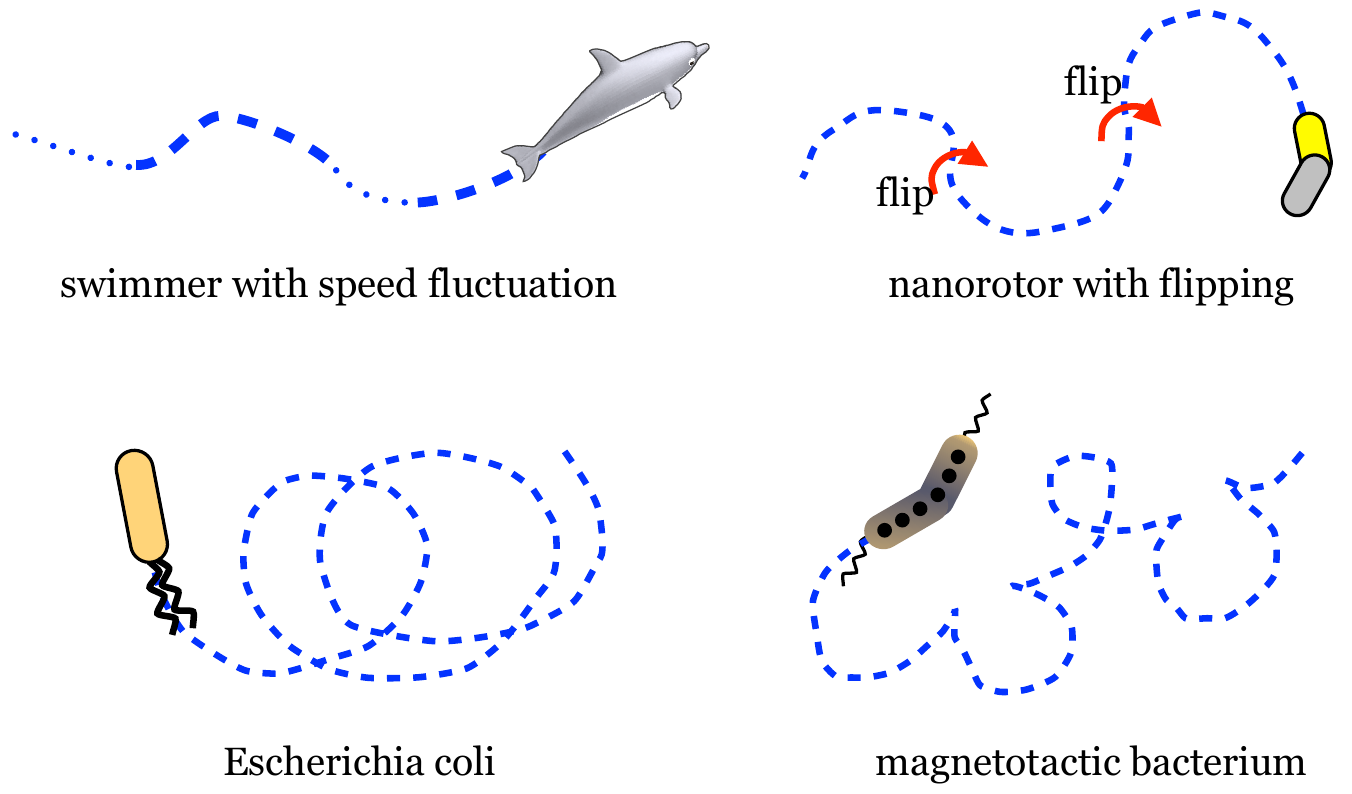}
\end{center}
\vspace{-15pt}
\caption{
The coupling of deterministic and stochastic elements can give rise to many
kinds of distinctive motion and spreading patterns, such as 
a rectilinear swimmer with speed fluctuation~\cite{Peruani:2007p602,Selmeczi:2008p603} and persistent turning~\cite{Gautrais2009JMB429,Degond2008JSP131},
a nanorotor with flipping~\cite{Takagi:2013p645}, 
E. coli circle swimming~\cite{
Lauga:2006p606,
PNAS-1995-Frymier-6195-9}
or 
magnetotactic bacteria with ocassional velocity reversal~\cite{
FLM:378272,                           
PhysRevE.73.021505,             
Erglis:2007p596,                      
Cebers:2011p639}.                  
}
\label{fig:spreading} 
\end{figure}

Self-propellers with active stochastic dynamics are motile non-equilibrium systems~\cite{Romanczuk:2012p624,
Marchetti2013RMP1143,
Chaudhuri2014PRE022131,
Lobaskin2008EPJST157}
ranging from 
bacteria~\cite{
Berg:2000p601,
Li-ProcNatlAcad:2008p614,
PNAS-1995-Frymier-6195-9,
Polin-Science:p487,
Saragosti:2012p636}, 
cells~\cite{
Sadati2014SBM137,
C3SM52893F,
Baker2014JRSI20140386,
Selmeczi:2008p603,
Selmeczi:2005p599,
Campos2010JTB526,
Li:2011p600,
Li:2008p594,
Haastert2004Nature626,
Riedel-Science:p300,
Friedrich:2008p610}, 
and 
nanomotors~\cite{
nourhani2013p062317,
Ebbens:2010p86,
Howse2007PRL048102,
Wang:2010p106,
Gibbs:2011p546}         
at the microscale to 
insects~\cite{
Theraulaz2002PNAS9645,
Casellas2008JTB424,
Jeanson2003JTB443}, 
fishes~\cite{
Niwa1994123,
Mach:2007p633,
Gautrais2009JMB429,
Degond2008JSP131},
and 
other animals~\cite{
Ordemann2003260,
citeulike:11429599,
Bazazi:2011p649,
Bazazi2008735,
Edwards:2007p598}, 
as well as
humans~\cite{RevModPhys.73.1067}
and 
traffic~\cite{Aw2002SIAM259}
 at the macroscale. 
The variety of stochastic fluctuations and their coupling with self-propellers' deterministic motion leads to distinct dynamical and spreading features (see Fig.~\ref{fig:spreading}).
Phenomenological modeling of self-propellers' ensemble behavior within the differential-equation based Langevin or Fokker-Planck formalisms grows mathematically cumbersome as the number of distinct elementary contributions to the dynamics grows. To overcome this difficulty, we recently described a kinematic matrix theory for self-propellers with uncorrelated (i.e.~white-noise) stochastic dynamics~\cite{Nourhani2014PRE062304}.
Here we advance this theory to include {\it correlated} Gaussian fluctuations -- colored noise. We demonstrate the formalism's utility by analyzing 
a rectilinear self-propeller with velocity fluctuations and orientational inertia and discuss the interplay of finite correlation times of the involved noises, leading to an emergent disorientation time scale and a variety of dynamical regimes.

The dynamics of a self-propeller can decompose into elementary processes such as deteministic translation and rotation as well as stochastic orientational diffusion, flips, and tumbles. In the white-noise limit of the kinematic matrix theory, the kinematic properties of these elementary processes are coded into a matrix, called the {\it kinematrix}, from which many ensemble properties of the self-propeller can be obtained by simple matrix algebra. This approach reveals universalities in self-propeller behavior that were previously hidden behind the complexity of differential-equation-based approaches~\cite{Nourhani2014PRE062304}. The approximation of a negligible stochastic correlation time has been used extensively to model self-propellers~\cite{
Romanczuk:2012p624,
Ebbens:2010p589,
Takagi:2013p645,
Nourhani2014PRE062304,
Nourhani2013p050301,
VanTeeffelen:2008p643,
Schweitzer1998PRL5044,
Cates:2013p620,
Romanczuk:2009p659,
Mandal2013PRL248101}. 
However, many physical systems suffer environmental noise in the form of forces that act directly on generalized momenta. Such noise is filtered through the {\em inertia} of the system and thus becomes colored~\cite{
Hanggi1995ACP239,
KlosekDygas1988SIAM425,
SanMiguel1980JSP605,
Kamenev2008PRL268103}. 
A similar picture holds for any system with a non-negligible response time, whether or not it fits into traditional mechanical descriptions. 
Invoking the Central Limit Theorem, we can reasonably expect a large fraction 
of such noises to be approximately Gaussian.

In this paper, first in section \ref{sec:theory} we advance the kinematrix approach to include Gaussian noise with finite correlation time, yielding Eqs. (\ref{eq:IntegroDiffGovEq}) -- (\ref{eq:GK}). This extended kinematrix formalism again circumvents the need for probability distributions; it also makes the calculations significantly easier by extracting the necessary information solely from the autocorrelation of the correlated noise. Section \ref{sec:linearmotor} then provides an application of the theory, to rectilinear self-propellers with fluctuating engines to study the effects of orientational Gaussian memory (modeled by an Ornstein-Uhlenbeck process) 
in producing a variety of ensemble regimes.

\section{Theory
\label{sec:theory}
}

The kinematrix formalism is based on an examination of elementary dynamical processes in a 
self-propeller's body frame. The tail-to-head vector ${\bm \chi}$ of the swimmer which has a
fixed orientation $\hat{\chi}$ in the body frame of the self-propeller evolves with time in the laboratory frame such that, for a given realization of noise, we have at time $\ubar{t}$ the updated value $\hat{\chi}(\ubar{t}) = U(0,\ubar{t}) \hat{\chi}(0)$. The propagator $\CU(0,\ubar{t})$ represents the net rotation of the body frame from 
time $0$ to time $\ubar{t}$; its ensemble average gives the velocity pair correlator of Eq.~(\ref{eq:velcorrelmem}) and the ensemble-average spatial displacement quantities of Eqs.~(\ref{eq:vectordisplace})--(\ref{eq:GK}).

To obtain $\langle U(0,\ubar{t}) \rangle$ we work on a discrete timeline
$\mathbb T = \{0,dt,2\, dt,3\, dt,\ldots\}$ with infinitesimal time steps $dt \ll t$.
We write $\CU_n$ for $\CU(0,n\, dt)$ and
$\CR_n$ for the net rotation between $n\, dt$ and $(n+1)\, dt$ in the laboratory frame;
in the body frame, the same rotation is expressed as 
$\tilde{\CR}_n = \CU_{n}^{-1} \CR_n \CU_{n}$. 
Rewriting the recursive expression $\CU_n = \CR_{n-1} \CU_{n-1}$ in terms of the body 
frame thus yields
\beq
\langle U_n \rangle = 
\langle \tilde{\CR}_0  \tilde{\CR}_1 \cdots  \tilde{\CR}_{n-1} \rangle
\label{eq:RawUt}
\eeq
where the brackets average over all possible realization of noises.
If the body-frame rotations $\tilde{\CR}_n$ are independent 
(the white-noise limit) then the average of their product in Eq.~(\ref{eq:RawUt}) 
is equal to the product of their averages. 
Then the expansion
$\langle \tilde{\CR}_i \rangle = {\cal I} - \ubar{\km} \, d\ubar{t} + 
{\cal O}(d\ubar{t}^{2})$ yields  
$\langle U(0,\ubar{t})\rangle=\exp(-\km \ubar{t})$ where the kinematrix
$\ubar{\km}$ captures the kinematic properties of the elementary motile processes~\cite{Nourhani2014PRE062304}.

However, for correlated noise the $\tilde{\CR}_n$'s are not independent. 
Assuming physically distinct and independent correlated and uncorrelated
noises, we write the rotation 
${\CR}_n = \Rcorr_n \Runcorr_n$ as the product of correlated $\Rcorr_n$ and uncorrelated $\Runcorr_n$ rotations (these being for an infinitesimal interval,
the ordering of the rotations makes a negligible difference).
Thus, $\langle U_{n+1} \rangle =  \langle U_n  \Rcorr_n \rangle \langle \Runcorr_n\rangle$.
The incremental correlated rotation can be written in the form
$\Rcorr_n= \exp(\swb_n \, d\ubar{t} \cdot \bm{\mathcal{J}})$, where 
the ${\mathcal{J}^\alpha}$ are the generators of rotations in $SO(3)$ (greek superscripts
denote Cartesian components $x$, $y$ and $z$). $\{\swb_n\}$ is assumed to comprise a stationary
centered Gaussian process with a continuous covariance:
$\langle \swb_n \rangle = 0$ and
$\langle \sw_n^\alpha \sw_m^\beta\rangle$ is a continuous function of $(n-m)dt$.
Expanding the exponential $\exp(\swb_n \, d\ubar{t} \cdot \bm{\mathcal{J}})$  
and expanding uncorrelated rotations to ${\cal O}(dt)$ as $\langle \tilde{\CR}_n^{\mbox{\scriptsize uncr}} \rangle \simeq {\cal I} - \ubar{\km}^{\mbox{\scriptsize uncr}} \, d\ubar{t}$  ($\ubar{\km}^{\mbox{\scriptsize uncr}}$ is the kinematrix of the uncorrelated elementary processes), we obtain
\begin{align}
\langle U_{n+1}\rangle 
= &\,
\langle U_n  \rangle -
\langle U_n  \rangle  \ubar{\km}^{\scriptstyle\text{uncr}} d\ubar{t} 
+
\left\langle U_n \left(\swb_n \cdot \bm{\mathcal{J}}\right)  \right\rangle d\ubar{t}
\nonumber \\
& 
+\mathcal{O}(dt^{2}).
\label{eq:differenceExpansion}
\end{align}
Large rotations are possible, but exceedingly rare. 
Their contribution to the expectation is negligible and we can work up to linear terms in $dt$.
Now, for a centered Gaussian-distributed vector $\bm{x}$ of any dimension,
the integration-by-parts identity
$\langle f(\bm{x}) x^\alpha \rangle = 
\sum_\beta \langle \partial f/\partial x^\beta \rangle \langle x^\beta x^\alpha\rangle$ holds~\cite{Brydges1982CMP123}.
Applying this identity and noting that $U_n$ depends only on $\swb_j$ for $j < n$ yields
\begin{equation*}
 \left\langle U_n  \swb_n 
\right\rangle \cdot \bm{\mathcal{J}}
\! = \! \,
\sum_{j < n}
\left[ \sum_{\alpha,\beta}
 \!
 \left\langle \sw_n^\alpha \sw_j^\beta \right \rangle
\! 
 \left\langle 
 \!
\tilde{\CR}_0  \cdots 
\!
 [{\cal J}_\beta \tilde{\CR}_j]
\!
 \cdots  \tilde {\CR}_n
\!
  \right\rangle \! {\cal J}_\alpha \right]
 d\ubar{t}.
\end{equation*}
Substituting into Eq.~(\ref{eq:differenceExpansion}) and 
reinterpreting the difference $(\langle U_{n+1}\rangle - \langle U_n \rangle)/dt$ 
as a derivative leads to
\begin{align}
{d \over d\ubar{t}} \langle U(0,\ubar{t}) \rangle
\!=\!&
\sum_{\alpha,\beta}\!
 \int_0^{\ubar{t}}\!\! \langle U(0,\ubar{t}^\prime) {\cal J}_\beta \, U(\ubar{t}^\prime,\ubar{t})\, {\cal J}_\alpha \rangle \!\left\langle \sw^\alpha(\ubar{t}) \sw^\beta(\ubar{t}^\prime) \right \rangle d\ubar{t}^\prime
\nonumber \\
&\! 
 - \langle U(0,\ubar{t})  \ubar\rangle{\km}^{\scriptstyle\text{uncr}}.
\label{eq:IntegroDiffGovEq}
\end{align}
The change of $\langle U(0,t)\rangle$ with time is due to the noise at time $t$.
Noise uncorrelated with what has gone before tends to degrade memory of
the past in a simple indiscriminate manner. But noise which is correlated
with the past, as $\swb$ is, has a more complicated effect.
Since a Gaussian distribution is determined by its mean
and covariance, the appearance of a simple covariance function 
in the governing equation~(\ref{eq:IntegroDiffGovEq}) is a natural consequence. 

Pretty as Eq.~(\ref{eq:IntegroDiffGovEq}) is, it becomes difficult to work with in
three or more dimensions since the matrices do not necessarily commute.
However, the two-dimensional case is already very rich and many
experimental studies involve self-propellers with an essentially two-dimensional
motion due to a confining planar substrate. 
We therefore confine ourselves to
the planar motion in the remainder of this paper.
All rotations are about the $z$ axis and 
the only matrices involved are
\beq
\mathcal{J}_z
=
\left[
 \begin{array}{ccc} 
  0 &  -1 & 0  \\
  1  &  0 & 0  \\ 
  0 &  0  &  0 
 \end{array} 
 \right],
\qquad
\mathcal{P}_z^\perp
=
\left[
 \begin{array}{ccc} 
  1 &  0 & 0  \\
  0 &  1 & 0  \\ 
  0 &  0  &  0 
 \end{array} 
\right]
 \label{eq:JzPz}
\eeq
where ${\cal J}_z$ is the generator of infinitesimal rotation about the $z$-axis and $\mathcal{P}_z^\perp$ projects into the $xy$-plane. Since $\langle U(0,\ubar{t}) \rangle$ is written in terms of $\ppz$ and $\jz$, and  $[\ppz,\jz]= \ppz\jz - \jz\ppz = 0$,
the commutation $[\langle U(0,\ubar{t}) \rangle, {\cal J}_z] = 0$ holds, and Eq.~(\ref{eq:IntegroDiffGovEq}) yields an exact solution in terms of the autocorrelation of the Gaussian noise:
\begin{subequations}
\begin{align}
&\langle U(0,\ubar{t}) \rangle = 
\exp\left[- \ubar{\km}^{\scriptstyle\text{uncr}} \ubar{t} - \ubar{\cal F}_\swb(\ubar{t}) \ppz \right]
\\
&
\ubar{\cal F}_\swb(\ubar{t}) =  {1 \over 2} \int_0^{\ubar{t}} \int_0^t\left\langle \sw(t^\prime) \sw(\ubar{t}^{\prime\prime}) \right \rangle d\ubar{t}^{\prime\prime} d\ubar{t}^\prime
\label{eq:genFxi}
\end{align}
\label{eq:2DGoverning}
\end{subequations}
By capturing the essential physics in the noise autocorrelation integral $\ubar{\cal F}_\swb(\ubar{t})$, the kinematrix treatment avoids the complication of dealing explicitly with  probability distributions 
by extracting the necessary information solely from the noise autocorrelation function.

A swimmer's tail-to-head direction $\hat{\chi}$ coincides with its instantaneous direction of deterministic velocity $\hat{v}$ in a rectilinear motion. While such a swimmer usually moves {\em forward} along tail-to-head axis ($\hat{v} = \hat{\chi}$), it can also occasionally swim {\em backward} along the same axis ($\hat{v} = - \hat{\chi}$).  
We reference the instantaneous velocity to the tail-to-head direction by writing
${\bm v} \equiv    v \hat{v} := v_s \hat{\chi}$. It is important to distinguish between $v$ and $v_s$ since the former is the speed (magnitude of the velocity) while the latter is a one dimensional velocity along the $\hat{\chi}$ axis such that for forward motion $v_s = v$ and for backward motion $v_s = -v$. As such, hereafter we refer to $v_s$ as ``signed-speed''.   

Now, choosing the laboratory frame such that $\unity \equiv \hat{\chi}(0)$, the velocity pair correlator $\langle {\bm v}(0)  \cdot {\bm v}({t}) \rangle$, the ensemble average of displacement $\langle \Delta  {\bm r}({t}) \rangle$, the mean square displacement $\langle|\Delta{\bm r}(t)|^2\rangle$, and effective diffusivity ${D}_{\scriptstyle\text{eff}}$ of the self-propeller can be obtained from
\begin{align}
&\langle {\bm v}(0)  \cdot   {\bm v}({t}) \rangle
 =
\langle v_s(0)  v_s(t) \rangle
\,
\langle U(0,{t})\rangle_{22},
 \label{eq:velcorrelmem}
\\
&
 \langle \Delta  {\bm r}({t}) \rangle  = \bar{v}_s \left[\int_0^{{t}}   \langle U(0,{t}^\prime)\rangle d{t}^\prime\right] \cdot \hat{\chi}(0),
\label{eq:vectordisplace}
 \\
&\langle|\Delta{\bm r}({t})|^2\rangle 
  =  
 2
 \int_0^{t} (t-t^\prime) \,
 \langle v_s(0)   v_s(t) \rangle
 \langle U(0,t^\prime)\rangle_{22}  d{t}^\prime,
 \label{eq:MSD1}
\end{align}
and
\beq
\Deff 
=
{1\over 2}  
\!
\int_0^\infty
\!
 \langle v_s(0)  v_s(t) \rangle
 \langle U(0,{t}^\prime)\rangle_{22}  dt^\prime,
\label{eq:GK} 
\eeq
where the subscript `22' denotes a matrix element. 
If there is no backwards motion (for example, the commonly-studied case of constant speed), then the propagator
can monitor the time evolution of the {\em dynamical} velocity vector rather than the {\em structural} tail-to-head vector. For such systems we make the modifications $v_s\mapsto v$ and $\hat{\chi} \mapsto \hat{v}$ in Eqs.~(\ref{eq:velcorrelmem})--(\ref{eq:GK}).

The main contribution of this paper is extending the kinematic matrix theory to include Gaussian memory, as expressed by Eqs.~(\ref{eq:IntegroDiffGovEq})--(\ref{eq:GK}). In the next section, as an example, we employ the formalism to discuss the physics of a rectilinear self-propeller with signed-speed fluctuation and Gaussian memory.

\section{Linear motion with fluctuating speed and Gaussian Memory
\label{sec:linearmotor}
} 

The interplay of multiple time scales of the elementary processes of motion determines
different regimes of swimmer ensemble behavior, quantified by asymptotic effective diffusivity and mean-square-displacement. 
In this section we study a self-propeller subjected to velocity fluctuations and 
orientational inertia, such as appears in the upper left of Fig.~\ref{fig:spreading}. 
Velocity fluctuations lead to stochastic variation of speed, which may also have inertial memory. The direction of motion may 
be influenced by stochastic noises arising from environmental fluctuations 
(e.g., 
Brownian kicks from fluid particles to a micron-sized self-propeller~\cite{Nourhani2013p050301,
Takagi:2013p645,
Ebbens:2010p589}, 
spatially scattered food supply~\cite{Strefler:2009p605}, 
or interaction with a substrate~\cite{Reichhardt2013PRE042306})  or internal fluctuations 
such as stochastic internal engine torque or decision-making processes of an organism.

Using a Fokker-Planck formalism, Peruani and Morelli~\cite{Peruani:2007p602} studied
a self-propeller with speed fluctuation and Brownian orientational diffusion, which
can account for internal engine fluctuations of biological systems.
However, the lack of orientational inertia cannot capture the essential physics of self-propeller dynamics
in many cases.
For instance, Gautrais {\it et al}.~\cite{Gautrais2009JMB429} analyzed 
trajectories of {\em Kuhlia mugil} fish swimming in a tank, observing constant speed 
motion with persistent turns that cannot be modeled  by a white noise.
Rather, there was an inertia associated with the angular velocity leading to a
decaying exponential autocorrelation. 
Correcting the white noise model with a finite inertial time leads to an Ornstein-Uhlenbeck 
process (OUP) for the angular velocity. Correspondingly, Gegond and Motsch~\cite{Degond2008JSP131} 
used a Fokker-Planck formalism to obtain the effective diffusivity of the fish with constant speed and OUP orientational dynamics.~Their model~\cite{Gautrais2009JMB429,Degond2008JSP131} matches
experimental data well, setting a solid ground for the presence of OUP dynamics in 
self-propeller dynamics.  
By adding a finite inertial time to a white noise, the OUP~\cite{OUoriginal1930} serves as the simplest colored 
noise that not only shows success in self-propellers
~\cite{
Weber:2011p644,
Gautrais2009JMB429,
Degond2008JSP131,
Dieterich2008PNAS459,
Radtke2012PRE051110,
Yushchenko2012PRE051127,
Szamel2014PRE012111,
Torney2008PRL078105}
, but also applies to other fields of 
physics such as 
quantum processes~\cite{
Olsen2013PRA051802,
Sarovar2012PRL130401,
Jing2010PRL40403,
Stimming02010PRL15301},
network dynamics~\cite{
Hu2014PRE032802}
and
genetics~\cite{
Assaf2013PRL058102,
Charlebois2011PRL218101,
Berg2008PRL188101}.

\begin{figure}[t]
\begin{center}
\includegraphics[width=2.8in]{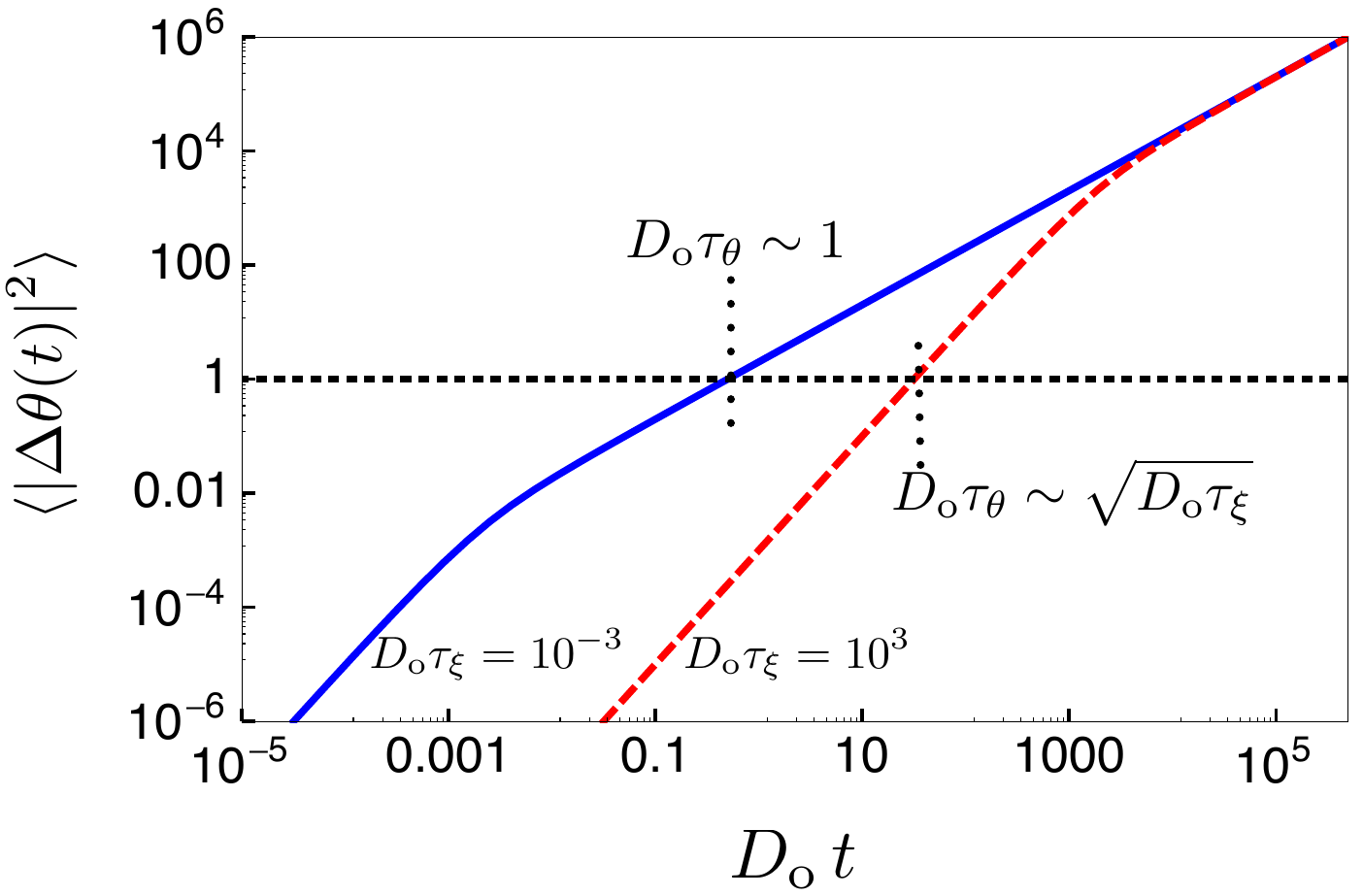}
\end{center}
\vspace{-15pt}
\caption{
The disorientation time for a self-propeller to forget its initial orientation depends on the ratio of inertial $\tauxi$ and orientational diffusion $\Dort^{-1}$ time scales. If the $\tauxi \ll \Dort$ we are close to the white noise limit and the self-propeller disorients over an orientational diffusion time scale $\tautheta\sim\Dort^{-1}$. On the other hand, when the inertial time is much larger than the orientational time scale, the disorientation time is the geometric average of inertial and orientational diffusion time scales, $\tautheta \sim (\tauxi \Dort^{-1})^{1/2}$.   
}
\label{fig:deltat2} 
\end{figure}

We analyze a more general model including both velocity fluctuations and orientational
inertia, subsuming the results of \cite{Degond2008JSP131,Peruani:2007p602}, yet 
with less complexity and more intuitive connection to the self-propeller physics.
The self-propeller moves in a plane at fluctuating velocity 
${\bm v}(t) = v_s \hat{\chi}$
and with an orientation $\theta$, defined by 
$\cos\theta = \unitx \cdot \hat{\chi}$.
The self-propeller's orientation changes 
according to 
\beq
{d\theta\over dt} = \xi,
\label{eq:angvel}
\eeq 
in which $\xi$ is a stationary 
OUP and $\eta$ is Gaussian white noise of intensity 
$\tauxi^{-2} \Dort$: 
\begin{subequations}
\begin{align}
&d\xi/dt = - \tauxi^{-1} {\xi}(t) + \eta(t)
\\
& \langle \eta(t) \eta(t^\prime) \rangle = 2 \tauxi^{-2} \Dort \delta(t-t^\prime),
\\
& \langle \sw(t) \sw(0)\rangle =\tauxi^{-1} \Dort e^{-|t|/\tauxi}
\label{eq:OUPCorrelation}
\end{align}
\label{eq:OUP}
\end{subequations}
Understanding the orientational wandering as being due to random torques, this model takes into account the self-propeller's rotational inertia. The variance of the angular
velocity, which may be a more convenient quantity for applications than $\Dort$,
is simply $\Dort/\tauxi$. 
In the limit $\tauxi\to 0$, $\sw$ acts as a white noise, recovering the simpler model of orientational Brownian motion diffusing at $\Dort$ with no inertia~\cite{Peruani:2007p602}. 
The autocorrelation integral [Eq.~(\ref{eq:genFxi})] for the OUP angular velocity $\xi$ is monotonically increasing:
\beq
 {\cal F}^{\mbox{\tiny OUP}} _\swb(t)
 = \Dort t +   \Dort t \left[{ e^{-t/\tauxi} -1 \over t/\tauxi}\right].
\label{eq:OUPFxi}
\eeq
The first term is the white-noise contribution and the second term is the modification due to inertia. 
Eqs.~(\ref{eq:genFxi}), (\ref{eq:angvel}) and (\ref{eq:OUPFxi}) yield
the mean square angular displacement 
\beq
\langle |\Delta\theta(t)|^2\rangle
=
2  {\cal F}^{\mbox{\tiny OUP}} _\swb(t)
\approx
 \begin{dcases}
        2\Dort t  &  t \gg \tauxi\\
         (t/\tauxi) \Dort t & t \ll \tauxi.
        \end{dcases}
 \label{eq:F-asymptotes}      
\eeq
Here, $\tauxi$ is the crossover time from ballistic to diffusive angular dynamics. However, we shall see below that the physical regime of the ensemble behavior is governed not only by $\tauxi$, but also the {\it disorientation time} $\tautheta$ over which the orientation changes significantly: $\langle |\Delta\theta(\tautheta)|^2\rangle \sim 1$. As illustrated in Fig.~(\ref{fig:deltat2}), $\tautheta$ can be distinct from both the orientational diffusion time $\Dort^{-1}$ and the inertial time $\tauxi$. If the inertial timescale is very short ($\tauxi \ll \Dort^{-1}$), then the self-propeller `forgets' its prior orientation through pure diffusion and $\tautheta \sim \Dort^{-1}$. If the inertial time is large ($\Dort^{-1} \ll \tauxi$), then $\langle |\Delta\theta|^2 \rangle$ becomes order one already in the ballistic regime and $\tautheta \sim (\Dort^{-1}\tauxi)^{1/2}$. Altogether, $\Dort\tautheta \sim \max\left(1, \sqrt{\Dort \tauxi}\right)$.
For example, the fish of \cite{Gautrais2009JMB429} have $\Dort\tauxi \sim 1/2$.

\begin{figure*}[t]
\begin{center}
\includegraphics[width=5.8in]{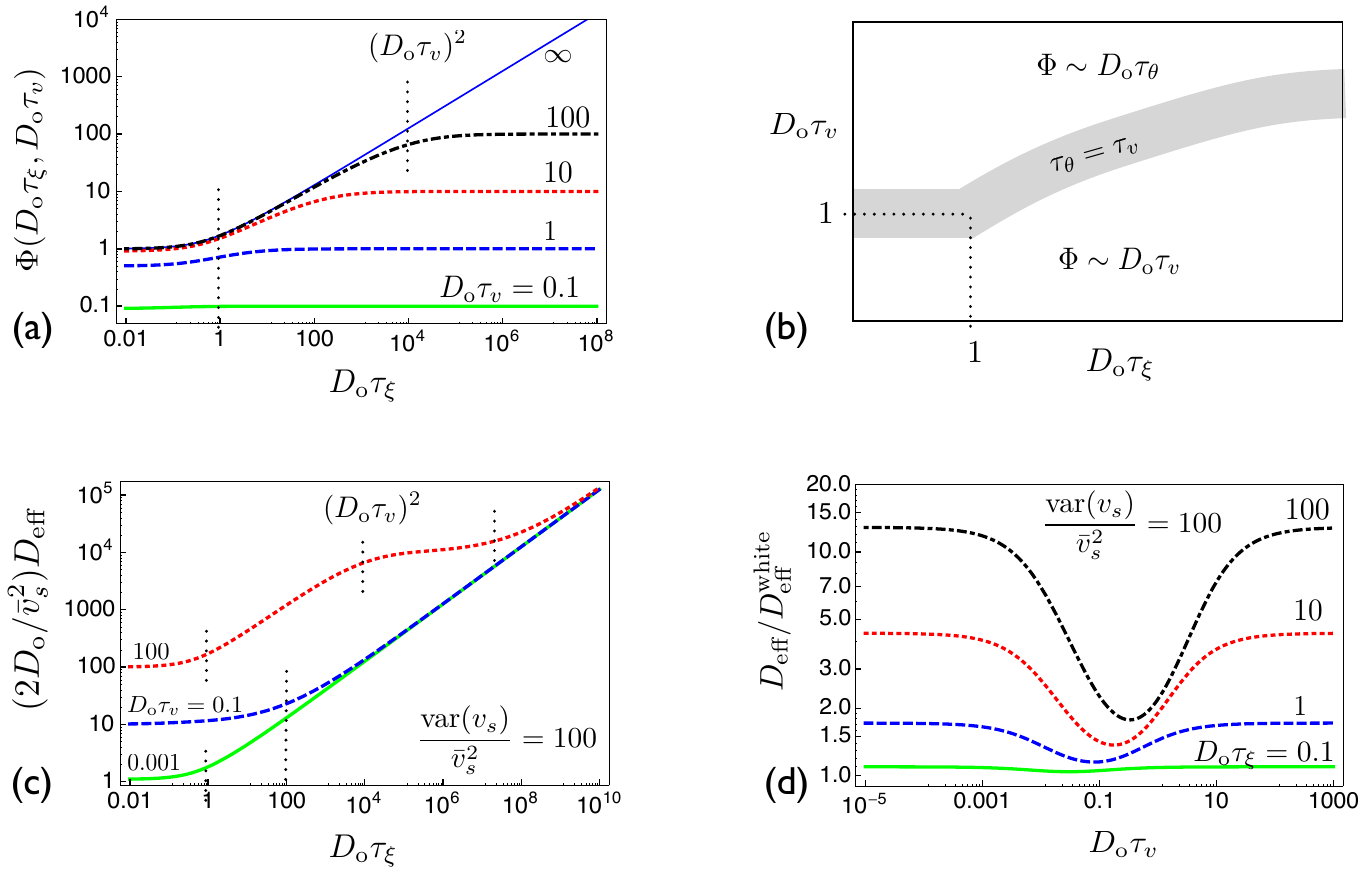}
\end{center}
\vspace{-15pt}
\caption{
Effective diffusion coefficient $\Deff$ of the linear self-propeller with orientational 
Gaussian noise characterized by correlation time $\tauxi$, speed fluctuations characterized by
correlation time $\tauv$ and asymptotic orientational diffusion coefficient $\Dort$.
(a) the mean-speed and fluctuation speed make contributions to $\Deff$ proportional
to $\Phi(\Dort\tauxi,\infty)$ and $\Phi(\Dort\tauxi,\Dort\tauv)$ respectively.
(b) $\Phi$ shows two major regimes depending upon whether the speed correlation time
is smaller or larger than the disorientation time $\tautheta$. In the former case,
the diffusion is essentially determined by speed fluctuations and in the latter
by orientational wandering. The disorientation time $\tautheta$ is proportional to
$(\Dort^{-1}\tauxi)^{1/2}$ when $\Dort \ll \tauxi$, and saturates to $\sim 1$ as
$\Dort\tauxi\to 0$. (c) $\Deff$ may contain several crossovers as a result of the
relative sizes and individual crossovers of the mean-speed and fluctuation-speed
components. Note that for $\Dort\tauxi \gg (\varv/\bar{v}_s^2)(\Dort\tauv)^2$, 
the mean-speed component always dominates. (d) Slices through $\Deff$ in the
other direction, normalized to the $\Dort\tauxi = 0$ value.
}
\label{fig:DeffMotor} 
\end{figure*}

Getting back to velocity fluctuations,
a signed-speed autocorrelation function  
\beq
\langle v_s(t) v_s(0) \rangle = 
\bar{v}_s^2  + \varv e^{-t/\tauv}.
\label{eq:speed-autocorrelation}
\eeq
appears naturally in many physical systems. 
It may arise from a self-propeller's interactions with the environment, varying terrain or fuel availability, and $\tauv$ reflects the inertia associated with signed-speed relaxation. 
The use of signed-speed subsumes the ordinary speed (velocity magnitude)
case where the motion is always
directed along the tail-to-head direction, but also situations where the motion
can sometimes be ``backward''. That might apply to crowded environments, such as 
for an individual cell in a cell monolayer
~\cite{
Sadati2014SBM137,
C3SM52893F,
Baker2014JRSI20140386,
Selmeczi:2008p603,
Selmeczi:2005p599,
Campos2010JTB526}.
If the dominance of forward over backward motion is slight, 
the dimensionless measure $\varv/\bar{v}_s^2$ of signed-speed fluctuations
can be very large. 
In that case, we observe multiple crossovers in the mean-square-displacement curves, as 
will be discussed later.  
The form (\ref{eq:speed-autocorrelation}) may represent a biased OUP processes with mean 
$\bar{v}_s$.
Alternatively, it may arise from internal engine fluctuations where the signed-speed jumps 
between discrete values. Such a case can be modeled by
a Poisson distribution (at rate $1/\tauv$) of ``reset times'' at each of which a new signed-speed is chosen independently from a fixed distribution with mean $\bar{v}_s$ and variance $\varv$. The path length between signed-speed resets has a mean $\bar{v}_s \tauv$ and variance $\varv \tau_v^2$.
For a self-propeller, a simple origin for such behavior might be a bistable engine,
giving two possible values for $v_s$. 

With the OUP autocorrelation integral~(\ref{eq:OUPFxi})  for persistent turning and the 
signed-speed autocorrelation function~(\ref{eq:speed-autocorrelation}) thus motivated,
we proceed to calculate the effective diffusivity $\Deff$ of the self-propeller using 
Eqs.~(\ref{eq:2DGoverning}) and (\ref{eq:GK}) as
\beq
\Deff 
 =
{\bar{v}_s^2 \over 2\Dort} {\Dcal}(\Dort\tau_\xi,\infty) 
 +  {\varv \over 2\Dort} \, {\Dcal}(\Dort\tau_\xi,\Dort\tau_v),
\label{eq:Deff-const+fluct-picture}
\eeq
where we have defined the dimensionless function $\Dcal$ with the following physical limits:
\begin{subequations}
\begin{align}
{\Dcal}(x,y) &\defeq
\!\!\!\!
\int_0^\infty 
\!
\exp
\!
\left\{\!- x\!\left[e^{-z/x}-1\right] - z\right\} e^{-{z}/{y}} \, dz
\label{eq:msfluccDeff}
\\
&= e^x  \sum_{k=0}^\infty {(-x)^k \big/ \,[k! \,(1 +  1/y  +  k/x)]}
\\[3pt]
& \approx
 \begin{dcases}
         \, e^x (1+1/y)^{-1}, & x \ll 1 
         \\[0.3em]
         (\mbox{$\frac{\pi}{2}$}\,x)^{1/2},             & 1 \ll x \ll y^2
         \\[0.3em]
         \, y,  & x \gg \max(1, y^2).
        \end{dcases}
 \label{eq:MDAD}  
\end{align}
\end{subequations}
The first term in the right-hand side of Eq.~(\ref{eq:Deff-const+fluct-picture}) describes the effective diffusion that would arise in the absence of signed-speed fluctuations, and the second term describes the unique contribution of signed-speed fluctuations to the effective diffusion. The reason for this clean separation is given below. 

Fig. \ref{fig:DeffMotor} plots $\Deff$ and $\Phi$ across a range of correlation times for orientation and speed. The diagram below facilitates an intuitive account of this behavior:
\begin{picture}(200,100)
\put(10,57){$\eta$:}
\put(20,57){$\mathrm{white}$}
\put(21,47){$\mathrm{noise}$}
\put(27,56){\circle{40}}
\put(50,52){\vector(1,0){35}}
\put(51,55){\text{inertia}}
\put(62,45){$\tau_\xi$}
\put(89,50){$\xi$}
\put(95,52){\vector(1,0){20}}
\put(118,49){$\theta$}
\put(127,56){\vector(2,1){27}}
\put(127,48){\vector(2,-1){27}}
\put(129,62){\rotatebox{30}{$\bar{v}_s$}}
\put(124,40){\rotatebox{-30}{$\varv$}}
\put(157,27){$\Delta{\bm r}^{\Fluct}$}
\put(140,15){\vector(2,1){15}}
\put(120,10){$\nu$,$\tauv$}
\put(128,12){\circle{21}}
\put(157,72){$\Delta{\bm r}^{\Const}$}
\put(175,67){\vector(2,-1){18}}
\put(177,37){\vector(2,1){16}}
\put(193,50){$\bigoplus$}
\put(205,53){\vector(1,0){12}}
\put(220,50){$\Delta{\bm r}$}
\end{picture}\\
The self-propeller's signed-speed can be split into a mean and a fluctuation: 
\beq
 v_s(t) = \bar{v}_s + \sqrt{\varv} \, \nu(t),
\eeq
where the noise $\nu$ obeys 
\beq
\langle \nu(t)\rangle = 0,\qquad \langle |\nu(t)|^2\rangle = 1.
\label{eq:normalizedfluc}
\eeq
The displacement can be similarly split as $\Delta{\bm r}(t) = \Delta{\bm r}^\Const(t) + \Delta{\bm r}^\Fluct(t)$. The diagram depicts the independent random inputs $\eta$ and $\nu$. Strictly speaking, $\Delta{\bm r}^\Const(t)$ and $\Delta{\bm r}^\Fluct(t)$ are not independent since they are driven by the same orientation process $\theta(t)$. But, 
they are {\em  probabilistically  orthogonal},
because the mean-speed and fluctuation-speed are: $\langle\nu(t)\rangle=0$. As a result, 
$\Delta{\bm r}^\Const(t)$ and $\Delta{\bm r}^\Fluct(t)$ (and through them
the mean signed-speed and signed-speed fluctuation) contribute to $\Deff$ in a simple additive way.

Three major features of  $\Phi(\Dort\tauxi,\Dort\tauv)$ in Fig.\ \ref{fig:DeffMotor}(a) leap to the eye. 
First, $\Phi(\Dort\tauxi,\infty)$, the curve for infinite $\Dort\tauv$ exhibits a 
crossover 
from a constant 1 to $\sim \sqrt{\Dort\tauxi}$ at 
$\Dort\tauxi \sim 1$. 
Second, for smaller values $1 \ll \Dort\tauv < \infty$, the curves follow that for 
$\Dort\tauv=\infty$ up to $\Dort \tauxi \sim (\Dort \tauv)^2$, at which
point they saturate to a value approximately $\Dort\tauv$.
Finally, for very small speed correlation time $\Dort\tauv \ll 1$, 
$\Phi(\Dort\tauxi,\Dort\tauv) \approx \Dort\tauv$ depends only weakly on
$\Dort\tauxi$.
An intuitive physical interpretation of these observations and the asymptotics in Eq.~(\ref{eq:MDAD}) follows from a comparison of the disorientation time $\tautheta$ and speed correlation time $\tauv$. Henceforth, we use a more precise definition for the disorientation time:
\beq
\Dort\tautheta \defeq \left[\max\left(1,\mbox{$\frac{\pi}{2}$}\Dort\tauxi \right)\right]^{1/2},
\label{eq:tau_theta}
\eeq
to recast Eq.(\ref{eq:MDAD}) into 
\beq
{\Dcal}(\Dort\tauxi,\Dort\tauv)
\approx
 \begin{dcases}
         \Dort\tauv, 
         & 
         \tau_v \ll \tautheta 
         \\[2pt]
         \Dort \tautheta,                     
         & 
         \tauv \gg \tautheta. \\
        \end{dcases}
 \label{eq:MDAD2}      
\eeq
Fig. \ref{fig:DeffMotor}(b) reveals two regimes of this equation, showing $\tautheta \sim \Dort^{-1}$ is independent of $\tauxi$ for $\Dort \tauxi \ll 1$. A straightforward understanding of (\ref{eq:MDAD2}) is at hand. 
To better understand this behavior, we rewrite
\begin{subequations}
\begin{align}
& {\bm \Gamma} : = \Delta {\bm r}^{\mbox{\scriptsize fluc}}/ [2 \Dort^{-1} \varv]^{1/2}
\\
& \Phi(\Dort\tauxi,\Dort\tauv) 
= 
\lim_{t\to\infty}{1 \over t}\langle|{\bm \Gamma}(t)|^2\rangle
\end{align}
\end{subequations}
and analyze $\Phi(\Dort\tauxi,\Dort\tauv)$ as the diffusive behavior of ${\bm \Gamma}$. In the limit $\tauv \ll \tautheta$ where signed-speed changes very rapidly
compared to orientation, the fluctuation part resembles a 
{\em one-dimensional} random walk along a slowly changing direction with step-duration $\Delta t = \tauv$ 
and step-length-squared $\langle |\Delta {\bm r}^{\mbox{\scriptsize fluc}}|^2 \rangle \approx  \langle |\nu(t)|^2\rangle \varv \tauv^2$.
By Eq.~(\ref{eq:normalizedfluc}),
$\Phi\approx \langle |\Delta {\bm \Gamma}|^2 \rangle/ \tauv \approx \Dort \tauv$.
In the opposite limit, $\tauv \gg \tautheta$, the fluctuation part has speed of order $\sqrt{\langle |\nu(t)|^2 \rangle \varv}$ which remains nearly constant during the time $\tautheta$, and resembles a two-dimensional random walker with step-duration $\tautheta$ and step-length-squared $\langle |\Delta {\bm r}^{\mbox{\scriptsize fluc}}|^2 \rangle = \varv \tautheta^2$. This leads to a $\nu$-averaged diffusivity $\Phi \approx \langle |{\bm \Gamma}|^2\rangle/\tautheta \approx \Dort \tautheta$. Fig. \ref{fig:DeffMotor}(a) now stands rationalized via Fig. \ref{fig:DeffMotor}(b) and Eq.~(\ref{eq:MDAD2}).

Turning now to the interpretation of the more complicated behavior of 
the effective diffusivity~(\ref{eq:Deff-const+fluct-picture}) depicted in Fig.~\ref{fig:DeffMotor}(c),
we note that the various asymptotic regimes can be collected into
\beq
2 \! \Deff 
\sim
{\bar{v}_s^2}{\tautheta}
 + {\varv} \min\left(\tauv, \tautheta\right).
\label{eq:general-approx-Deff} 
\eeq
The critical parameter determining the number of crossovers is $\Dort\tauv$. 

If the orientational diffusion time scale greatly exceeds the speed correlation time $(\tauv \ll \Dort^{-1})$, we have $ \min\left(\tauv, \tautheta\right) = \tauv$;
the fluctuation-speed contribution $\varv\tauv$ is independent of $\Dort\tauxi$ and 
the only question is when this dominates the mean-speed contribution.
In case $[\varv/\bar{v}^2] \Dort\tauv \ll 1$, the answer is never. This is exemplified
by the solid green hockey-stick shaped curve in Fig.~\ref{fig:DeffMotor}(c).
Otherwise, there is a crossover from fluctuation-speed domination to mean-speed
domination at 
$\Dort \tauxi \simeq [{\varv \over \bar{v}_s^2}\Dort\tauv]^2$, as shown in the blue dashed curve.

On the other hand, if the time required for changing signed-speed is much longer than the orientational diffusion time scale ($\tauv \gg \Dort^{-1}$),
the story starts off similarly with a roughly constant value
$2\Deff \approx [\bar{v}_s^2 + \varv]\Dort^{-1}$ up to about
$\tauxi \sim \Dort^{-1}$, at which point it shifts into
the arm of the hockey stick with $2\Deff \approx[\bar{v}_s^2 + \varv](\Dort^{-1}\tauxi)^{1/2}$. 
But, when $\Dort\tauxi$ exceeds $(\Dort\tauv)^2$, the mean-speed contribution
continues to increase, while the fluctuation-speed contribution plateaus. 
If $\varv/\bar{v}_s^2 \gg 1$, this appears as a clear 
plateau, as seen in the dotted red curve of Fig.~\ref{fig:DeffMotor}(c),
until the mean-speed contribution
becomes dominant at $\tauxi \sim \Dort [{\varv \over \bar{v}_s^2}\tauv]^2$
and the $\sqrt{\Dort\tauxi}$ behavior of the hockey stick returns.

\begin{figure}[t]
\begin{center}
\includegraphics[width=2.7in]{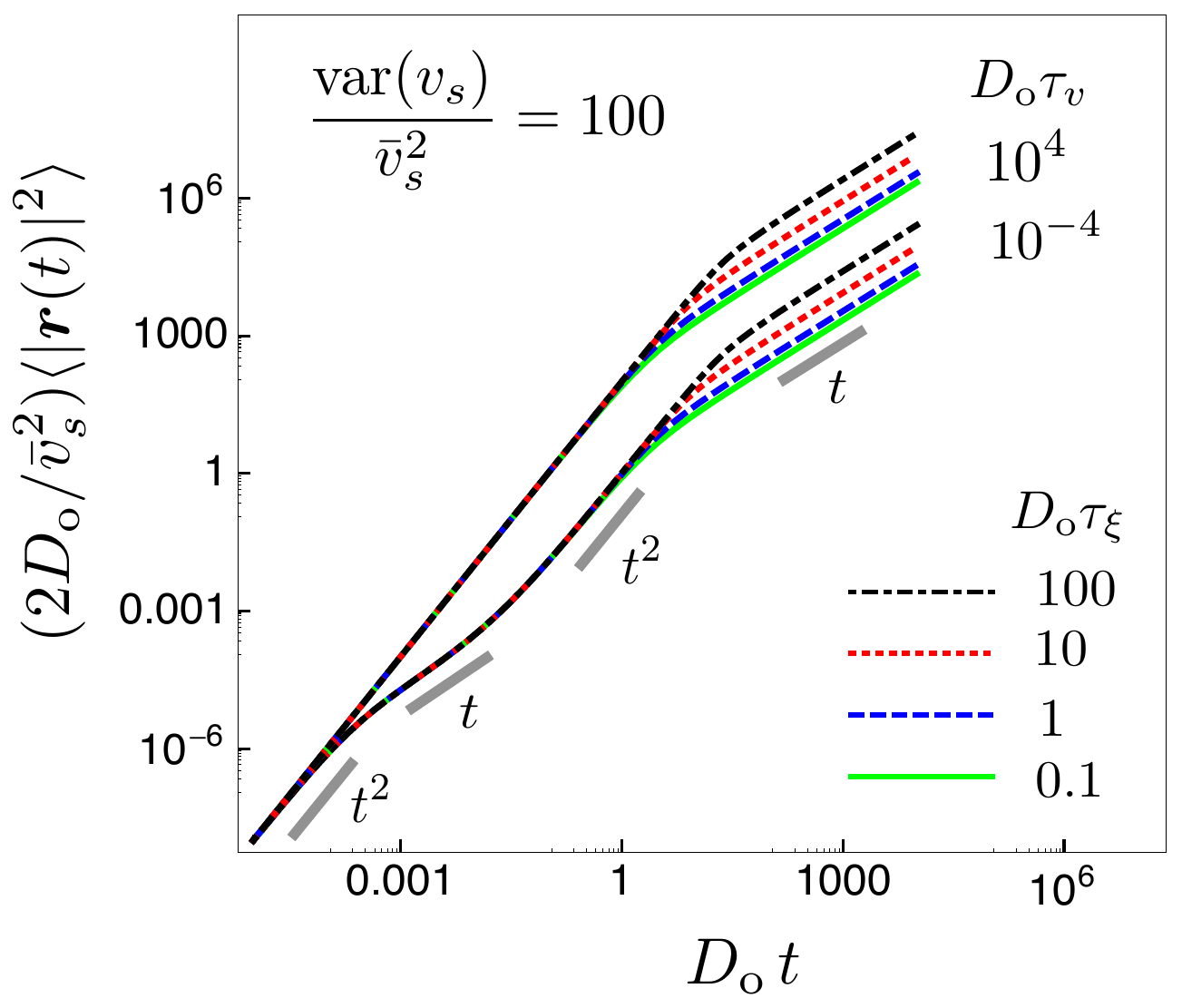}
\end{center}
\vspace{-15pt}
\caption{
Mean-squared displacement of the linear self-propeller with Gaussian orientational 
noise of correlation time $\tauxi$, signed-speed fluctuations with correlation time
$\tauv$, and asymptotic orientational diffusion coefficient $\Dort$.
If the self-propeller disorients faster than signed-speed changes value, $\tautheta \ll \tauv$, (upper curves) we observe a single crossover from ballistic to diffusive 
regimes at $\tautheta$; with increase in inertial time $\tauxi$ the crossover happens 
later. In the opposite regime $\tautheta \gg \tauv$ (lower curves) for large 
signed-speed fluctuations $\varv/\bar{v}_s^2 \gg 1$ the signed-speed fluctuation 
contribution shows a one-dimensional ballistic to diffusive crossover about 
$\tauv$. Combined with mean signed-speed part 
(behaving like the $\tautheta \ll \tauv$ case), this produces three crossovers.
}
\label{fig:mu106crossover} 
\end{figure}

Fig. \ref{fig:DeffMotor}(d) gives another perspective on $\Deff$ by slicing in the
other direction and taking a ratio to the white noise limit ($\Dort \tauxi\to0$)
\beq
{D}_{\scriptstyle\text{eff}}^{\scriptstyle\text{white}} = \frac{\bar{v}_s^2}{2\Dort} 
 + \frac{\varv}{2\Dort}{1\over  (1+1/\Dort\tauv)}.
\label{eq:Deff-white}
\eeq
In the limit $\Dort\tauv \to 0$  of rapid speed fluctuations 
the effective diffusivity depends only on average speed
 $\Deff \sim {\bar{v}_s^2}{\tautheta}$, and as $\Dort\tauv \to \infty$, $\Deff \sim [\bar{v}_s^2+\varv]\tautheta$. In either extreme, $\Deff$ is simply proportional to $\tautheta$, 
which is $\Dort^{-1}$ in the white noise limit.
 So, for {\em both} very large and very small $\Dort\tauv$,
$\Deff/{D}_{\scriptstyle\text{eff}}^{\scriptstyle\text{white}} \approx \Dort\tautheta$,
 independently of the signed-speed parameters $\bar{v}_s$ and $\varv$. 
In between, if $\varv/\bar{v}_s^2$ is large enough, there is a region where
$\Deff$ is insensitive to $\Dort\tauxi$ up to a large value. This corresponds
to the long dashed blue plateau shown at $\Dort\tauv = 0.1$ in Fig.~\ref{fig:DeffMotor}(c).

The effective diffusivity characterizes only the asymptotic behavior of the 
mean-square displacement. The full time dependence exhibits additional complexity.
Using Eq.~(\ref{eq:MSD1}) we obtain the mean square displacement
\begin{align}
\langle|\Delta {\bm r}(t)|^2\rangle \defeq 
& \, 4t \Deff + 4 t \, \frac{\bar{v}_s^2}{2\Dort} \, \tilde{\Dcal}(\Dort\tauxi,\infty,\Dort t) 
\nonumber \\
&+ 4 t \, \frac{\varv}{2\Dort} \,  \tilde{\Dcal}(\Dort\tauxi,\Dort\tauv,\Dort t),
\label{eq:MSDLinearMotor}
\end{align}
where
\begin{subequations}
\begin{align}
\tilde{\Dcal}(x,y,z) &=
{e^x \over z}  \sum_{k=0}^\infty {(-x)^k 
\left[ 
e^{-(1 + 1/y + k/x) z}-1
\right]
\over k!  (1 + 1/y + k/x)^2 } 
\\[3pt]
&\approx
 \begin{dcases}
        \mbox{${1\over 2}$}z - \Dcal(x,y), & z\ll\min(1,x,y) \\
        0,  & z\to\infty. \\[3pt]
        \end{dcases}
\end{align}
\end{subequations}
For times much shorter than all the characteristic time scales, 
$t \ll \min(\Dort^{-1},\tauxi,\tauv)$, we have ballistic motion 
$\langle|\Delta {\bm r}(t)|^2\rangle \approx [\bar{v}_s^2 + \varv] t^2$, independently of orientational, inertial and signed-speed correlation time scales. At very long times the self-propeller behaves 
diffusively;
$\langle|\Delta {\bm r}(t\to\infty)|^2\rangle$ is  $4 t \Deff$ and depends on 
all three time scales $\Dort^{-1}$, $\tauxi$ and $\tauv$.
Fig.~\ref{fig:mu106crossover} shows the behavior of Eq.~(\ref{eq:MSDLinearMotor}) in the limit of 
large speed fluctuations for a variety of time scales. 
Since $\langle |\Delta{\bm r}^\Const(t)|^2\rangle$ and $\langle |\Delta{\bm r}^\Fluct(t)|^2\rangle$ each has its own ballistic-to-diffusive  crossover,
in the limit of rapid, large signed-speed fluctuations, i.e. $\Dort\tauv\ll1$ and $\varv/\bar{v}_s^2 \gg 1$, three clear crossovers are observed. 
 We analyze the two limiting regimes $\tauv \ll \tautheta$ and $\tautheta \ll \tauv$. 
The mean-speed contribution $\langle |\Delta{\bm r}^\Const(t)|^2\rangle$ is invariably in the latter limit.
Suppose first that the self-propeller disorients much faster than its signed-speed changes, i.e., $\tautheta \ll \tauv$.
 Then, an individual self-propeller has a ballistic-to-diffusive crossover at time $\tautheta$. Its speed is stable over much longer times, so it behaves as though its diffusion coefficient were fluctuating on the time scale $\tauv$. On the other hand, if $\tauv \ll \tautheta$, 
the fluctuation part of the displacement [$\Delta{\bm r}^\Fluct(t)$] 
 of an individual self-propeller has a ballistic-to-diffusive crossover at $\tauv$, but to a {\em nearly one-dimensional} diffusive motion
since much before the self-propeller disorients, the signed-speed has changed many times. There is a second crossover, to genuinely two-dimensional diffusion, at $\tautheta$ 
when the self-propeller starts to disorient.
 In either case, however, the ensemble average $\langle |\Delta{\bm r}^\Fluct(t)|^2\rangle$ will evidence only the primary crossover at $\min(\tautheta,\tauv)$.
When $\tautheta \ll \tauv$ (the upper set of curves in Fig. \ref{fig:mu106crossover}), the full mean-square displacement exhibits just a single ballistic-to-diffusive crossover at $\tautheta$.
In case $\tauv \ll \tautheta$, the speed-fluctuation contribution 
$\Delta{\bm r}^\Fluct$ becomes diffusive earlier.
If $\langle|\Delta{\bm r}^\Const(\tautheta)|^2\rangle/\tautheta \sim
\bar{v}_s^2 \tautheta \gg \varv \tauv
\sim \langle|\Delta{\bm r}^\Fluct(\tautheta)|^2\rangle/\tautheta$, 
the total motion can re-enter a 
ballistic regime when $\Delta{\bm r}^\Const$ comes to dominate somewhere
between $\tauv$ and $\tautheta$. Later, at $\tautheta$,
this component, too, becomes diffusive. This is exemplified by the 
lower set of curves in Fig. \ref{fig:mu106crossover}.

\section{concluding remarks}
The extension of kinematic matrix theory to incorporate correlated Gaussian noises expands its 
applicability to real-world systems with significant inertia.
The ability to work straightforwardly from just the noise autocorrelation 
simplifies calculations significantly and helps one to focus more on the physics of 
the problem.
Our streamlined and close-to-the-physics treatment of
the rectilinear self-propeller with velocity fluctuations and persistent turning ---
a model with real-world interest~\cite{Peruani:2007p602,Degond2008JSP131} ---
exemplifies this. 
This simplicity of kinematic matrix theory enables the study of more complicated 
systems with less mathematical sophistication, and provides a useful tool for experimentalists 
to develop models for analyzing their data. 

The general governing equation (\ref{eq:IntegroDiffGovEq}) applies also in 
higher dimensions where the simplifying feature of commutation of all the
rotations disappears. Concrete development of the kinematrix approach 
to self-propellers moving in three dimensions will be a natural and useful 
direction for further study.

\section*{Acknowledgements}
This work was supported by the NSF under Grant No. DMR-0820404 through the Penn State Center for Nanoscale Science.


\begin{thebibliography}{76}%
\makeatletter
\providecommand \@ifxundefined [1]{%
 \@ifx{#1\undefined}
}%
\providecommand \@ifnum [1]{%
 \ifnum #1\expandafter \@firstoftwo
 \else \expandafter \@secondoftwo
 \fi
}%
\providecommand \@ifx [1]{%
 \ifx #1\expandafter \@firstoftwo
 \else \expandafter \@secondoftwo
 \fi
}%
\providecommand \natexlab [1]{#1}%
\providecommand \enquote  [1]{``#1''}%
\providecommand \bibnamefont  [1]{#1}%
\providecommand \bibfnamefont [1]{#1}%
\providecommand \citenamefont [1]{#1}%
\providecommand \href@noop [0]{\@secondoftwo}%
\providecommand \href [0]{\begingroup \@sanitize@url \@href}%
\providecommand \@href[1]{\@@startlink{#1}\@@href}%
\providecommand \@@href[1]{\endgroup#1\@@endlink}%
\providecommand \@sanitize@url [0]{\catcode `\\12\catcode `\$12\catcode
  `\&12\catcode `\#12\catcode `\^12\catcode `\_12\catcode `\%12\relax}%
\providecommand \@@startlink[1]{}%
\providecommand \@@endlink[0]{}%
\providecommand \url  [0]{\begingroup\@sanitize@url \@url }%
\providecommand \@url [1]{\endgroup\@href {#1}{\urlprefix }}%
\providecommand \urlprefix  [0]{URL }%
\providecommand \Eprint [0]{\href }%
\providecommand \doibase [0]{http://dx.doi.org/}%
\providecommand \selectlanguage [0]{\@gobble}%
\providecommand \bibinfo  [0]{\@secondoftwo}%
\providecommand \bibfield  [0]{\@secondoftwo}%
\providecommand \translation [1]{[#1]}%
\providecommand \BibitemOpen [0]{}%
\providecommand \bibitemStop [0]{}%
\providecommand \bibitemNoStop [0]{.\EOS\space}%
\providecommand \EOS [0]{\spacefactor3000\relax}%
\providecommand \BibitemShut  [1]{\csname bibitem#1\endcsname}%
\let\auto@bib@innerbib\@empty
\bibitem [{\citenamefont {Romanczuk}\ \emph {et~al.}(2012)\citenamefont
  {Romanczuk}, \citenamefont {B{\"a}r}, \citenamefont {Ebeling}, \citenamefont
  {Lindner},\ and\ \citenamefont {Schimansky-Geier}}]{Romanczuk:2012p624}%
  \BibitemOpen
  \bibfield  {author} {\bibinfo {author} {\bibfnamefont {P.}~\bibnamefont
  {Romanczuk}}, \bibinfo {author} {\bibfnamefont {M.}~\bibnamefont {B{\"a}r}},
  \bibinfo {author} {\bibfnamefont {W.}~\bibnamefont {Ebeling}}, \bibinfo
  {author} {\bibfnamefont {B.}~\bibnamefont {Lindner}}, \ and\ \bibinfo
  {author} {\bibfnamefont {L.}~\bibnamefont {Schimansky-Geier}},\ }\href
  {\doibase 10.1140/epjst/e2012-01529-y} {\bibfield  {journal} {\bibinfo
  {journal} {Eur. Phys. J. Spec. Top.}\ }\textbf {\bibinfo {volume} {202}},\
  \bibinfo {pages} {1} (\bibinfo {year} {2012})}\BibitemShut {NoStop}%
\bibitem [{\citenamefont {Marchetti}\ \emph {et~al.}(2013)\citenamefont
  {Marchetti}, \citenamefont {Joanny}, \citenamefont {Ramaswamy}, \citenamefont
  {Liverpool}, \citenamefont {Prost}, \citenamefont {Rao},\ and\ \citenamefont
  {Simha}}]{Marchetti2013RMP1143}%
  \BibitemOpen
  \bibfield  {author} {\bibinfo {author} {\bibfnamefont {M.~C.}\ \bibnamefont
  {Marchetti}}, \bibinfo {author} {\bibfnamefont {J.~F.}\ \bibnamefont
  {Joanny}}, \bibinfo {author} {\bibfnamefont {S.}~\bibnamefont {Ramaswamy}},
  \bibinfo {author} {\bibfnamefont {T.~B.}\ \bibnamefont {Liverpool}}, \bibinfo
  {author} {\bibfnamefont {J.}~\bibnamefont {Prost}}, \bibinfo {author}
  {\bibfnamefont {M.}~\bibnamefont {Rao}}, \ and\ \bibinfo {author}
  {\bibfnamefont {R.~A.}\ \bibnamefont {Simha}},\ }\href@noop {} {\bibfield
  {journal} {\bibinfo  {journal} {Rev. Mod. Phys.}\ }\textbf {\bibinfo {volume}
  {85}},\ \bibinfo {pages} {1143} (\bibinfo {year} {2013})}\BibitemShut
  {NoStop}%
\bibitem [{\citenamefont {Chaudhuri}(2014)}]{Chaudhuri2014PRE022131}%
  \BibitemOpen
  \bibfield  {author} {\bibinfo {author} {\bibfnamefont {D.}~\bibnamefont
  {Chaudhuri}},\ }\href@noop {} {\bibfield  {journal} {\bibinfo  {journal}
  {Phys. Rev. E}\ }\textbf {\bibinfo {volume} {90}},\ \bibinfo {pages} {022131}
  (\bibinfo {year} {2014})}\BibitemShut {NoStop}%
\bibitem [{\citenamefont {Lobaskin}\ \emph {et~al.}(2008)\citenamefont
  {Lobaskin}, \citenamefont {Lobaskin},\ and\ \citenamefont
  {Kulic}}]{Lobaskin2008EPJST157}%
  \BibitemOpen
  \bibfield  {author} {\bibinfo {author} {\bibfnamefont {V.}~\bibnamefont
  {Lobaskin}}, \bibinfo {author} {\bibfnamefont {D.}~\bibnamefont {Lobaskin}},
  \ and\ \bibinfo {author} {\bibfnamefont {I.~M.}\ \bibnamefont {Kulic}},\
  }\href@noop {} {\bibfield  {journal} {\bibinfo  {journal} {Eur. Phys. J.
  Spec. Top.}\ }\textbf {\bibinfo {volume} {157}},\ \bibinfo {pages} {149}
  (\bibinfo {year} {2008})}\BibitemShut {NoStop}%
\bibitem [{\citenamefont {Berg}(2000)}]{Berg:2000p601}%
  \BibitemOpen
  \bibfield  {author} {\bibinfo {author} {\bibfnamefont {H.~C.}\ \bibnamefont
  {Berg}},\ }\href@noop {} {\bibfield  {journal} {\bibinfo  {journal} {Physics
  Today}\ }\textbf {\bibinfo {volume} {53}},\ \bibinfo {pages} {24} (\bibinfo
  {year} {2000})}\BibitemShut {NoStop}%
\bibitem [{\citenamefont {Li}\ \emph {et~al.}(2008{\natexlab{a}})\citenamefont
  {Li}, \citenamefont {Tam},\ and\ \citenamefont
  {Tang}}]{Li-ProcNatlAcad:2008p614}%
  \BibitemOpen
  \bibfield  {author} {\bibinfo {author} {\bibfnamefont {G.}~\bibnamefont
  {Li}}, \bibinfo {author} {\bibfnamefont {L.-K.}\ \bibnamefont {Tam}}, \ and\
  \bibinfo {author} {\bibfnamefont {J.}~\bibnamefont {Tang}},\ }\href
  {http://www.pnas.org/content/105/47/18355.short} {\bibfield  {journal}
  {\bibinfo  {journal} {Proc. Natl. Acad. Sci. USA}\ }\textbf {\bibinfo
  {volume} {105}},\ \bibinfo {pages} {18355} (\bibinfo {year}
  {2008}{\natexlab{a}})}\BibitemShut {NoStop}%
\bibitem [{\citenamefont {Frymier}\ \emph {et~al.}(1995)\citenamefont
  {Frymier}, \citenamefont {Ford}, \citenamefont {Berg},\ and\ \citenamefont
  {Cummings}}]{PNAS-1995-Frymier-6195-9}%
  \BibitemOpen
  \bibfield  {author} {\bibinfo {author} {\bibfnamefont {P.~D.}\ \bibnamefont
  {Frymier}}, \bibinfo {author} {\bibfnamefont {R.~M.}\ \bibnamefont {Ford}},
  \bibinfo {author} {\bibfnamefont {H.~C.}\ \bibnamefont {Berg}}, \ and\
  \bibinfo {author} {\bibfnamefont {P.~T.}\ \bibnamefont {Cummings}},\
  }\href@noop {} {\bibfield  {journal} {\bibinfo  {journal} {Proc. Natl. Acad.
  Sci. USA}\ }\textbf {\bibinfo {volume} {92}},\ \bibinfo {pages} {6195 }
  (\bibinfo {year} {1995})}\BibitemShut {NoStop}%
\bibitem [{\citenamefont {Polin}\ \emph {et~al.}(2009)\citenamefont {Polin},
  \citenamefont {Tuval}, \citenamefont {Drescher}, \citenamefont {Gollub},\
  and\ \citenamefont {Goldstein}}]{Polin-Science:p487}%
  \BibitemOpen
  \bibfield  {author} {\bibinfo {author} {\bibfnamefont {M.}~\bibnamefont
  {Polin}}, \bibinfo {author} {\bibfnamefont {I.}~\bibnamefont {Tuval}},
  \bibinfo {author} {\bibfnamefont {K.}~\bibnamefont {Drescher}}, \bibinfo
  {author} {\bibfnamefont {J.~P.}\ \bibnamefont {Gollub}}, \ and\ \bibinfo
  {author} {\bibfnamefont {R.~E.}\ \bibnamefont {Goldstein}},\ }\href@noop {}
  {\bibfield  {journal} {\bibinfo  {journal} {Science}\ }\textbf {\bibinfo
  {volume} {325}},\ \bibinfo {pages} {487} (\bibinfo {year}
  {2009})}\BibitemShut {NoStop}%
\bibitem [{\citenamefont {Saragosti}\ \emph {et~al.}(2012)\citenamefont
  {Saragosti}, \citenamefont {Silberzan},\ and\ \citenamefont
  {Buguin}}]{Saragosti:2012p636}%
  \BibitemOpen
  \bibfield  {author} {\bibinfo {author} {\bibfnamefont {J.}~\bibnamefont
  {Saragosti}}, \bibinfo {author} {\bibfnamefont {P.}~\bibnamefont
  {Silberzan}}, \ and\ \bibinfo {author} {\bibfnamefont {A.}~\bibnamefont
  {Buguin}},\ }\href {\doibase 10.1371/journal.pone.0035412.s001} {\bibfield
  {journal} {\bibinfo  {journal} {PLoS ONE}\ }\textbf {\bibinfo {volume} {7}},\
  \bibinfo {pages} {e35412} (\bibinfo {year} {2012})}\BibitemShut {NoStop}%
\bibitem [{\citenamefont {Sadati}\ \emph {et~al.}(2014)\citenamefont {Sadati},
  \citenamefont {Nourhani}, \citenamefont {Fredberg},\ and\ \citenamefont
  {Taheri~Qazvini}}]{Sadati2014SBM137}%
  \BibitemOpen
  \bibfield  {author} {\bibinfo {author} {\bibfnamefont {M.}~\bibnamefont
  {Sadati}}, \bibinfo {author} {\bibfnamefont {A.}~\bibnamefont {Nourhani}},
  \bibinfo {author} {\bibfnamefont {J.~J.}\ \bibnamefont {Fredberg}}, \ and\
  \bibinfo {author} {\bibfnamefont {N.}~\bibnamefont {Taheri~Qazvini}},\
  }\href@noop {} {\bibfield  {journal} {\bibinfo  {journal} {WIREs Syst. Biol.
  Med.}\ }\textbf {\bibinfo {volume} {6}},\ \bibinfo {pages} {137} (\bibinfo
  {year} {2014})}\BibitemShut {NoStop}%
\bibitem [{\citenamefont {Bi}\ \emph {et~al.}(2014)\citenamefont {Bi},
  \citenamefont {Lopez}, \citenamefont {Schwarz},\ and\ \citenamefont
  {Manning}}]{C3SM52893F}%
  \BibitemOpen
  \bibfield  {author} {\bibinfo {author} {\bibfnamefont {D.}~\bibnamefont
  {Bi}}, \bibinfo {author} {\bibfnamefont {J.~H.}\ \bibnamefont {Lopez}},
  \bibinfo {author} {\bibfnamefont {J.~M.}\ \bibnamefont {Schwarz}}, \ and\
  \bibinfo {author} {\bibfnamefont {M.~L.}\ \bibnamefont {Manning}},\ }\href
  {\doibase 10.1039/C3SM52893F} {\bibfield  {journal} {\bibinfo  {journal}
  {Soft Matter}\ }\textbf {\bibinfo {volume} {10}},\ \bibinfo {pages} {1885}
  (\bibinfo {year} {2014})}\BibitemShut {NoStop}%
\bibitem [{\citenamefont {Baker}\ \emph {et~al.}(2014)\citenamefont {Baker},
  \citenamefont {Brasch}, \citenamefont {Manning},\ and\ \citenamefont
  {Henderson}}]{Baker2014JRSI20140386}%
  \BibitemOpen
  \bibfield  {author} {\bibinfo {author} {\bibfnamefont {R.~M.}\ \bibnamefont
  {Baker}}, \bibinfo {author} {\bibfnamefont {M.~E.}\ \bibnamefont {Brasch}},
  \bibinfo {author} {\bibfnamefont {M.~L.}\ \bibnamefont {Manning}}, \ and\
  \bibinfo {author} {\bibfnamefont {J.~H.}\ \bibnamefont {Henderson}},\
  }\href@noop {} {\bibfield  {journal} {\bibinfo  {journal} {J. R. Soc.
  Interface}\ }\textbf {\bibinfo {volume} {11}},\ \bibinfo {pages} {20140386}
  (\bibinfo {year} {2014})}\BibitemShut {NoStop}%
\bibitem [{\citenamefont {Selmeczi}\ \emph {et~al.}(2008)\citenamefont
  {Selmeczi}, \citenamefont {Li}, \citenamefont {Pedersen}, \citenamefont
  {Nrrelykke}, \citenamefont {Hagedorn}, \citenamefont {Mosler}, \citenamefont
  {Larsen}, \citenamefont {Cox},\ and\ \citenamefont
  {Flyvbjerg}}]{Selmeczi:2008p603}%
  \BibitemOpen
  \bibfield  {author} {\bibinfo {author} {\bibfnamefont {D.}~\bibnamefont
  {Selmeczi}}, \bibinfo {author} {\bibfnamefont {L.}~\bibnamefont {Li}},
  \bibinfo {author} {\bibfnamefont {L.~I.}\ \bibnamefont {Pedersen}}, \bibinfo
  {author} {\bibfnamefont {S.~F.}\ \bibnamefont {Nrrelykke}}, \bibinfo {author}
  {\bibfnamefont {P.~H.}\ \bibnamefont {Hagedorn}}, \bibinfo {author}
  {\bibfnamefont {S.}~\bibnamefont {Mosler}}, \bibinfo {author} {\bibfnamefont
  {N.~B.}\ \bibnamefont {Larsen}}, \bibinfo {author} {\bibfnamefont {E.~C.}\
  \bibnamefont {Cox}}, \ and\ \bibinfo {author} {\bibfnamefont
  {H.}~\bibnamefont {Flyvbjerg}},\ }\href {\doibase
  10.1140/epjst/e2008-00626-x} {\bibfield  {journal} {\bibinfo  {journal} {Eur.
  Phys. J. Spec. Top.}\ }\textbf {\bibinfo {volume} {157}},\ \bibinfo {pages}
  {1} (\bibinfo {year} {2008})}\BibitemShut {NoStop}%
\bibitem [{\citenamefont {Selmeczi}\ \emph {et~al.}(2005)\citenamefont
  {Selmeczi}, \citenamefont {Mosler}, \citenamefont {Hagedorn}, \citenamefont
  {Larsen},\ and\ \citenamefont {Flyvbjerg}}]{Selmeczi:2005p599}%
  \BibitemOpen
  \bibfield  {author} {\bibinfo {author} {\bibfnamefont {D.}~\bibnamefont
  {Selmeczi}}, \bibinfo {author} {\bibfnamefont {S.}~\bibnamefont {Mosler}},
  \bibinfo {author} {\bibfnamefont {P.~H.}\ \bibnamefont {Hagedorn}}, \bibinfo
  {author} {\bibfnamefont {N.~B.}\ \bibnamefont {Larsen}}, \ and\ \bibinfo
  {author} {\bibfnamefont {H.}~\bibnamefont {Flyvbjerg}},\ }\href {\doibase
  10.1529/biophysj.105.061150} {\bibfield  {journal} {\bibinfo  {journal}
  {Biophys. J.}\ }\textbf {\bibinfo {volume} {89}},\ \bibinfo {pages} {912}
  (\bibinfo {year} {2005})}\BibitemShut {NoStop}%
\bibitem [{\citenamefont {Campos}\ \emph {et~al.}(2010)\citenamefont {Campos},
  \citenamefont {Mendez},\ and\ \citenamefont {Llopis}}]{Campos2010JTB526}%
  \BibitemOpen
  \bibfield  {author} {\bibinfo {author} {\bibfnamefont {D.}~\bibnamefont
  {Campos}}, \bibinfo {author} {\bibfnamefont {V.}~\bibnamefont {Mendez}}, \
  and\ \bibinfo {author} {\bibfnamefont {I.}~\bibnamefont {Llopis}},\
  }\href@noop {} {\bibfield  {journal} {\bibinfo  {journal} {J. Theor. Biol.}\
  }\textbf {\bibinfo {volume} {267}},\ \bibinfo {pages} {526} (\bibinfo {year}
  {2010})}\BibitemShut {NoStop}%
\bibitem [{\citenamefont {Li}\ \emph {et~al.}(2011)\citenamefont {Li},
  \citenamefont {Cox},\ and\ \citenamefont {Flyvbjerg}}]{Li:2011p600}%
  \BibitemOpen
  \bibfield  {author} {\bibinfo {author} {\bibfnamefont {L.}~\bibnamefont
  {Li}}, \bibinfo {author} {\bibfnamefont {E.~C.}\ \bibnamefont {Cox}}, \ and\
  \bibinfo {author} {\bibfnamefont {H.}~\bibnamefont {Flyvbjerg}},\ }\href
  {\doibase 10.1088/1478-3975/8/4/046006} {\bibfield  {journal} {\bibinfo
  {journal} {Phys. Biol.}\ }\textbf {\bibinfo {volume} {8}},\ \bibinfo {pages}
  {046006} (\bibinfo {year} {2011})}\BibitemShut {NoStop}%
\bibitem [{\citenamefont {Li}\ \emph {et~al.}(2008{\natexlab{b}})\citenamefont
  {Li}, \citenamefont {N{\o}rrelykke},\ and\ \citenamefont
  {Cox}}]{Li:2008p594}%
  \BibitemOpen
  \bibfield  {author} {\bibinfo {author} {\bibfnamefont {L.}~\bibnamefont
  {Li}}, \bibinfo {author} {\bibfnamefont {S.~F.}\ \bibnamefont
  {N{\o}rrelykke}}, \ and\ \bibinfo {author} {\bibfnamefont {E.~C.}\
  \bibnamefont {Cox}},\ }\href {\doibase 10.1371/journal.pone.0002093.s004}
  {\bibfield  {journal} {\bibinfo  {journal} {PLoS ONE}\ }\textbf {\bibinfo
  {volume} {3}},\ \bibinfo {pages} {e2093} (\bibinfo {year}
  {2008}{\natexlab{b}})}\BibitemShut {NoStop}%
\bibitem [{\citenamefont {Van~Haastert}\ and\ \citenamefont
  {Devreotes}(2004)}]{Haastert2004Nature626}%
  \BibitemOpen
  \bibfield  {author} {\bibinfo {author} {\bibfnamefont {P.~J.~M.}\
  \bibnamefont {Van~Haastert}}\ and\ \bibinfo {author} {\bibfnamefont {P.~N.}\
  \bibnamefont {Devreotes}},\ }\href@noop {} {\bibfield  {journal} {\bibinfo
  {journal} {Nature Reviews Molecular Cell Biology}\ }\textbf {\bibinfo
  {volume} {5}},\ \bibinfo {pages} {626} (\bibinfo {year} {2004})}\BibitemShut
  {NoStop}%
\bibitem [{\citenamefont {Riedel}\ \emph {et~al.}(2005)\citenamefont {Riedel},
  \citenamefont {Kruse},\ and\ \citenamefont {Howard}}]{Riedel-Science:p300}%
  \BibitemOpen
  \bibfield  {author} {\bibinfo {author} {\bibfnamefont {I.~H.}\ \bibnamefont
  {Riedel}}, \bibinfo {author} {\bibfnamefont {K.}~\bibnamefont {Kruse}}, \
  and\ \bibinfo {author} {\bibfnamefont {J.}~\bibnamefont {Howard}},\
  }\href@noop {} {\bibfield  {journal} {\bibinfo  {journal} {Science}\ }\textbf
  {\bibinfo {volume} {309}},\ \bibinfo {pages} {300} (\bibinfo {year}
  {2005})}\BibitemShut {NoStop}%
\bibitem [{\citenamefont {Friedrich}\ and\ \citenamefont
  {J{\"u}licher}(2008)}]{Friedrich:2008p610}%
  \BibitemOpen
  \bibfield  {author} {\bibinfo {author} {\bibfnamefont {B.~M.}\ \bibnamefont
  {Friedrich}}\ and\ \bibinfo {author} {\bibfnamefont {F.}~\bibnamefont
  {J{\"u}licher}},\ }\href {\doibase 10.1088/1367-2630/10/12/123025} {\bibfield
   {journal} {\bibinfo  {journal} {New J. Phys.}\ }\textbf {\bibinfo {volume}
  {10}},\ \bibinfo {pages} {123025} (\bibinfo {year} {2008})}\BibitemShut
  {NoStop}%
\bibitem [{\citenamefont {Nourhani}\ \emph
  {et~al.}(2013{\natexlab{a}})\citenamefont {Nourhani}, \citenamefont {Byun},
  \citenamefont {Lammert}, \citenamefont {Borhan},\ and\ \citenamefont
  {Crespi}}]{nourhani2013p062317}%
  \BibitemOpen
  \bibfield  {author} {\bibinfo {author} {\bibfnamefont {A.}~\bibnamefont
  {Nourhani}}, \bibinfo {author} {\bibfnamefont {Y.-M.}\ \bibnamefont {Byun}},
  \bibinfo {author} {\bibfnamefont {P.~E.}\ \bibnamefont {Lammert}}, \bibinfo
  {author} {\bibfnamefont {A.}~\bibnamefont {Borhan}}, \ and\ \bibinfo {author}
  {\bibfnamefont {V.~H.}\ \bibnamefont {Crespi}},\ }\href@noop {} {\bibfield
  {journal} {\bibinfo  {journal} {Phys. Rev. E}\ }\textbf {\bibinfo {volume}
  {88}},\ \bibinfo {pages} {062317} (\bibinfo {year}
  {2013}{\natexlab{a}})}\BibitemShut {NoStop}%
\bibitem [{\citenamefont {Ebbens}\ and\ \citenamefont
  {Howse}(2010)}]{Ebbens:2010p86}%
  \BibitemOpen
  \bibfield  {author} {\bibinfo {author} {\bibfnamefont {S.~J.}\ \bibnamefont
  {Ebbens}}\ and\ \bibinfo {author} {\bibfnamefont {J.~R.}\ \bibnamefont
  {Howse}},\ }\href {\doibase 10.1039/b918598d} {\bibfield  {journal} {\bibinfo
   {journal} {Soft Matter}\ }\textbf {\bibinfo {volume} {6}},\ \bibinfo {pages}
  {726} (\bibinfo {year} {2010})}\BibitemShut {NoStop}%
\bibitem [{\citenamefont {Howse}\ \emph {et~al.}(2007)\citenamefont {Howse},
  \citenamefont {Jones}, \citenamefont {Ryan}, \citenamefont {Gough},
  \citenamefont {Vafabakhsh},\ and\ \citenamefont
  {Golestanian}}]{Howse2007PRL048102}%
  \BibitemOpen
  \bibfield  {author} {\bibinfo {author} {\bibfnamefont {J.~R.}\ \bibnamefont
  {Howse}}, \bibinfo {author} {\bibfnamefont {R.~A.~L.}\ \bibnamefont {Jones}},
  \bibinfo {author} {\bibfnamefont {A.~J.}\ \bibnamefont {Ryan}}, \bibinfo
  {author} {\bibfnamefont {T.}~\bibnamefont {Gough}}, \bibinfo {author}
  {\bibfnamefont {R.}~\bibnamefont {Vafabakhsh}}, \ and\ \bibinfo {author}
  {\bibfnamefont {R.}~\bibnamefont {Golestanian}},\ }\href@noop {} {\bibfield
  {journal} {\bibinfo  {journal} {Phys. Rev. Lett.}\ }\textbf {\bibinfo
  {volume} {99}},\ \bibinfo {pages} {048102} (\bibinfo {year}
  {2007})}\BibitemShut {NoStop}%
\bibitem [{\citenamefont {Wang}\ and\ \citenamefont
  {Manesh}(2010)}]{Wang:2010p106}%
  \BibitemOpen
  \bibfield  {author} {\bibinfo {author} {\bibfnamefont {J.}~\bibnamefont
  {Wang}}\ and\ \bibinfo {author} {\bibfnamefont {K.~M.}\ \bibnamefont
  {Manesh}},\ }\href {\doibase 10.1002/smll.200901746} {\bibfield  {journal}
  {\bibinfo  {journal} {Small}\ }\textbf {\bibinfo {volume} {6}},\ \bibinfo
  {pages} {338} (\bibinfo {year} {2010})}\BibitemShut {NoStop}%
\bibitem [{\citenamefont {Gibbs}\ and\ \citenamefont
  {Zhao}(2011)}]{Gibbs:2011p546}%
  \BibitemOpen
  \bibfield  {author} {\bibinfo {author} {\bibfnamefont {J.}~\bibnamefont
  {Gibbs}}\ and\ \bibinfo {author} {\bibfnamefont {Y.}~\bibnamefont {Zhao}},\
  }\href {http://www.springerlink.com/index/X023V857423481XG.pdf} {\bibfield
  {journal} {\bibinfo  {journal} {Front. Mater. Sci.}\ }\textbf {\bibinfo
  {volume} {5}},\ \bibinfo {pages} {25} (\bibinfo {year} {2011})}\BibitemShut
  {NoStop}%
\bibitem [{\citenamefont {Theraulaz}\ \emph {et~al.}(2002)\citenamefont
  {Theraulaz}, \citenamefont {Bonabeau}, \citenamefont {Nicolis}, \citenamefont
  {Sole}, \citenamefont {Fourcassie}, \citenamefont {Blanco}, \citenamefont
  {Fournier}, \citenamefont {Joly}, \citenamefont {Fernandez}, \citenamefont
  {Grimal}, \citenamefont {Dalle},\ and\ \citenamefont
  {Deneubourg}}]{Theraulaz2002PNAS9645}%
  \BibitemOpen
  \bibfield  {author} {\bibinfo {author} {\bibfnamefont {G.}~\bibnamefont
  {Theraulaz}}, \bibinfo {author} {\bibfnamefont {E.}~\bibnamefont {Bonabeau}},
  \bibinfo {author} {\bibfnamefont {S.~C.}\ \bibnamefont {Nicolis}}, \bibinfo
  {author} {\bibfnamefont {R.~V.}\ \bibnamefont {Sole}}, \bibinfo {author}
  {\bibfnamefont {V.}~\bibnamefont {Fourcassie}}, \bibinfo {author}
  {\bibfnamefont {S.}~\bibnamefont {Blanco}}, \bibinfo {author} {\bibfnamefont
  {R.}~\bibnamefont {Fournier}}, \bibinfo {author} {\bibfnamefont {J.-L.}\
  \bibnamefont {Joly}}, \bibinfo {author} {\bibfnamefont {P.}~\bibnamefont
  {Fernandez}}, \bibinfo {author} {\bibfnamefont {A.}~\bibnamefont {Grimal}},
  \bibinfo {author} {\bibfnamefont {P.}~\bibnamefont {Dalle}}, \ and\ \bibinfo
  {author} {\bibfnamefont {J.-L.}\ \bibnamefont {Deneubourg}},\ }\href@noop {}
  {\bibfield  {journal} {\bibinfo  {journal} {P Natl Acad Sci Usa}\ }\textbf
  {\bibinfo {volume} {99}},\ \bibinfo {pages} {9645} (\bibinfo {year}
  {2002})}\BibitemShut {NoStop}%
\bibitem [{\citenamefont {Casellas}\ \emph {et~al.}(2008)\citenamefont
  {Casellas}, \citenamefont {Gautrais}, \citenamefont {Fournier}, \citenamefont
  {Blanco}, \citenamefont {Combe}, \citenamefont {Fourcassie}, \citenamefont
  {Theraulaz},\ and\ \citenamefont {Jost}}]{Casellas2008JTB424}%
  \BibitemOpen
  \bibfield  {author} {\bibinfo {author} {\bibfnamefont {E.}~\bibnamefont
  {Casellas}}, \bibinfo {author} {\bibfnamefont {J.}~\bibnamefont {Gautrais}},
  \bibinfo {author} {\bibfnamefont {R.}~\bibnamefont {Fournier}}, \bibinfo
  {author} {\bibfnamefont {S.}~\bibnamefont {Blanco}}, \bibinfo {author}
  {\bibfnamefont {M.}~\bibnamefont {Combe}}, \bibinfo {author} {\bibfnamefont
  {V.}~\bibnamefont {Fourcassie}}, \bibinfo {author} {\bibfnamefont
  {G.}~\bibnamefont {Theraulaz}}, \ and\ \bibinfo {author} {\bibfnamefont
  {C.}~\bibnamefont {Jost}},\ }\href@noop {} {\bibfield  {journal} {\bibinfo
  {journal} {J. Theor. Biol.}\ }\textbf {\bibinfo {volume} {250}},\ \bibinfo
  {pages} {424} (\bibinfo {year} {2008})}\BibitemShut {NoStop}%
\bibitem [{\citenamefont {Jeanson}\ \emph {et~al.}(2003)\citenamefont
  {Jeanson}, \citenamefont {Blanco}, \citenamefont {Fournier}, \citenamefont
  {Deneubourg}, \citenamefont {Fourcassi{\'e}},\ and\ \citenamefont
  {Theraulaz}}]{Jeanson2003JTB443}%
  \BibitemOpen
  \bibfield  {author} {\bibinfo {author} {\bibfnamefont {R.}~\bibnamefont
  {Jeanson}}, \bibinfo {author} {\bibfnamefont {S.}~\bibnamefont {Blanco}},
  \bibinfo {author} {\bibfnamefont {R.}~\bibnamefont {Fournier}}, \bibinfo
  {author} {\bibfnamefont {J.~L.}\ \bibnamefont {Deneubourg}}, \bibinfo
  {author} {\bibfnamefont {V.}~\bibnamefont {Fourcassi{\'e}}}, \ and\ \bibinfo
  {author} {\bibfnamefont {G.}~\bibnamefont {Theraulaz}},\ }\href@noop {}
  {\bibfield  {journal} {\bibinfo  {journal} {J. Theor. Biol.}\ }\textbf
  {\bibinfo {volume} {225}},\ \bibinfo {pages} {443} (\bibinfo {year}
  {2003})}\BibitemShut {NoStop}%
\bibitem [{\citenamefont {Niwa}(1994)}]{Niwa1994123}%
  \BibitemOpen
  \bibfield  {author} {\bibinfo {author} {\bibfnamefont {H.-S.}\ \bibnamefont
  {Niwa}},\ }\href {\doibase http://dx.doi.org/10.1006/jtbi.1994.1218}
  {\bibfield  {journal} {\bibinfo  {journal} {J. Theor. Biol.}\ }\textbf
  {\bibinfo {volume} {171}},\ \bibinfo {pages} {123 } (\bibinfo {year}
  {1994})}\BibitemShut {NoStop}%
\bibitem [{\citenamefont {Mach}\ and\ \citenamefont
  {Schweitzer}(2007)}]{Mach:2007p633}%
  \BibitemOpen
  \bibfield  {author} {\bibinfo {author} {\bibfnamefont {R.}~\bibnamefont
  {Mach}}\ and\ \bibinfo {author} {\bibfnamefont {F.}~\bibnamefont
  {Schweitzer}},\ }\href {\doibase 10.1007/s11538-006-9135-3} {\bibfield
  {journal} {\bibinfo  {journal} {Bull. Math. Biol.}\ }\textbf {\bibinfo
  {volume} {69}},\ \bibinfo {pages} {539} (\bibinfo {year} {2007})}\BibitemShut
  {NoStop}%
\bibitem [{\citenamefont {Gautrais}\ \emph {et~al.}(2009)\citenamefont
  {Gautrais}, \citenamefont {Jost}, \citenamefont {Soria}, \citenamefont
  {Campo}, \citenamefont {Motsch}, \citenamefont {Fournier}, \citenamefont
  {Blanco},\ and\ \citenamefont {Theraulaz}}]{Gautrais2009JMB429}%
  \BibitemOpen
  \bibfield  {author} {\bibinfo {author} {\bibfnamefont {J.}~\bibnamefont
  {Gautrais}}, \bibinfo {author} {\bibfnamefont {C.}~\bibnamefont {Jost}},
  \bibinfo {author} {\bibfnamefont {M.}~\bibnamefont {Soria}}, \bibinfo
  {author} {\bibfnamefont {A.}~\bibnamefont {Campo}}, \bibinfo {author}
  {\bibfnamefont {S.}~\bibnamefont {Motsch}}, \bibinfo {author} {\bibfnamefont
  {R.}~\bibnamefont {Fournier}}, \bibinfo {author} {\bibfnamefont
  {S.}~\bibnamefont {Blanco}}, \ and\ \bibinfo {author} {\bibfnamefont
  {G.}~\bibnamefont {Theraulaz}},\ }\href@noop {} {\bibfield  {journal}
  {\bibinfo  {journal} {J. Math. Biol.}\ }\textbf {\bibinfo {volume} {58}},\
  \bibinfo {pages} {429} (\bibinfo {year} {2009})}\BibitemShut {NoStop}%
\bibitem [{\citenamefont {Degond}\ and\ \citenamefont
  {Motsch}(2008)}]{Degond2008JSP131}%
  \BibitemOpen
  \bibfield  {author} {\bibinfo {author} {\bibfnamefont {P.}~\bibnamefont
  {Degond}}\ and\ \bibinfo {author} {\bibfnamefont {S.}~\bibnamefont
  {Motsch}},\ }\href@noop {} {\bibfield  {journal} {\bibinfo  {journal} {J.
  Stat. Phys.}\ }\textbf {\bibinfo {volume} {131}},\ \bibinfo {pages} {989}
  (\bibinfo {year} {2008})}\BibitemShut {NoStop}%
\bibitem [{\citenamefont {Ordemann}\ \emph {et~al.}(2003)\citenamefont
  {Ordemann}, \citenamefont {Balazsi},\ and\ \citenamefont
  {Moss}}]{Ordemann2003260}%
  \BibitemOpen
  \bibfield  {author} {\bibinfo {author} {\bibfnamefont {A.}~\bibnamefont
  {Ordemann}}, \bibinfo {author} {\bibfnamefont {G.}~\bibnamefont {Balazsi}}, \
  and\ \bibinfo {author} {\bibfnamefont {F.}~\bibnamefont {Moss}},\ }\href
  {\doibase http://dx.doi.org/10.1016/S0378-4371(03)00204-8} {\bibfield
  {journal} {\bibinfo  {journal} {Physica A: Statistical Mechanics and its
  Applications}\ }\textbf {\bibinfo {volume} {325}},\ \bibinfo {pages} {260 }
  (\bibinfo {year} {2003})}\BibitemShut {NoStop}%
\bibitem [{\citenamefont {Komin}\ \emph {et~al.}(2004)\citenamefont {Komin},
  \citenamefont {Erdmann},\ and\ \citenamefont
  {Schimansky-Geier}}]{citeulike:11429599}%
  \BibitemOpen
  \bibfield  {author} {\bibinfo {author} {\bibfnamefont {N.}~\bibnamefont
  {Komin}}, \bibinfo {author} {\bibfnamefont {U.}~\bibnamefont {Erdmann}}, \
  and\ \bibinfo {author} {\bibfnamefont {L.}~\bibnamefont {Schimansky-Geier}},\
  }\href {\doibase 10.1142/s0219477504001756} {\bibfield  {journal} {\bibinfo
  {journal} {Fluc. Noise Lett}\ }\textbf {\bibinfo {volume} {4}},\ \bibinfo
  {pages} {L151} (\bibinfo {year} {2004})}\BibitemShut {NoStop}%
\bibitem [{\citenamefont {Bazazi}\ \emph {et~al.}(2011)\citenamefont {Bazazi},
  \citenamefont {Romanczuk}, \citenamefont {Thomas}, \citenamefont
  {Schimansky-Geier}, \citenamefont {Hale}, \citenamefont {Miller},
  \citenamefont {Sword}, \citenamefont {Simpson},\ and\ \citenamefont
  {Couzin}}]{Bazazi:2011p649}%
  \BibitemOpen
  \bibfield  {author} {\bibinfo {author} {\bibfnamefont {S.}~\bibnamefont
  {Bazazi}}, \bibinfo {author} {\bibfnamefont {P.}~\bibnamefont {Romanczuk}},
  \bibinfo {author} {\bibfnamefont {S.}~\bibnamefont {Thomas}}, \bibinfo
  {author} {\bibfnamefont {L.}~\bibnamefont {Schimansky-Geier}}, \bibinfo
  {author} {\bibfnamefont {J.~J.}\ \bibnamefont {Hale}}, \bibinfo {author}
  {\bibfnamefont {G.~A.}\ \bibnamefont {Miller}}, \bibinfo {author}
  {\bibfnamefont {G.~A.}\ \bibnamefont {Sword}}, \bibinfo {author}
  {\bibfnamefont {S.~J.}\ \bibnamefont {Simpson}}, \ and\ \bibinfo {author}
  {\bibfnamefont {I.~D.}\ \bibnamefont {Couzin}},\ }\href {\doibase
  10.1098/rspb.1993.0044} {\bibfield  {journal} {\bibinfo  {journal} {Proc.
  Roy. Soc. B: Bio. Sci.}\ }\textbf {\bibinfo {volume} {278}},\ \bibinfo
  {pages} {356} (\bibinfo {year} {2011})}\BibitemShut {NoStop}%
\bibitem [{\citenamefont {Bazazi}\ \emph {et~al.}(2008)\citenamefont {Bazazi},
  \citenamefont {Buhl}, \citenamefont {Hale}, \citenamefont {Anstey},
  \citenamefont {Sword}, \citenamefont {Simpson},\ and\ \citenamefont
  {Couzin}}]{Bazazi2008735}%
  \BibitemOpen
  \bibfield  {author} {\bibinfo {author} {\bibfnamefont {S.}~\bibnamefont
  {Bazazi}}, \bibinfo {author} {\bibfnamefont {J.}~\bibnamefont {Buhl}},
  \bibinfo {author} {\bibfnamefont {J.~J.}\ \bibnamefont {Hale}}, \bibinfo
  {author} {\bibfnamefont {M.~L.}\ \bibnamefont {Anstey}}, \bibinfo {author}
  {\bibfnamefont {G.~A.}\ \bibnamefont {Sword}}, \bibinfo {author}
  {\bibfnamefont {S.~J.}\ \bibnamefont {Simpson}}, \ and\ \bibinfo {author}
  {\bibfnamefont {I.~D.}\ \bibnamefont {Couzin}},\ }\href {\doibase
  http://dx.doi.org/10.1016/j.cub.2008.04.035} {\bibfield  {journal} {\bibinfo
  {journal} {Current Biology}\ }\textbf {\bibinfo {volume} {18}},\ \bibinfo
  {pages} {735 } (\bibinfo {year} {2008})}\BibitemShut {NoStop}%
\bibitem [{\citenamefont {Edwards}\ \emph {et~al.}(2007)\citenamefont
  {Edwards}, \citenamefont {Phillips}, \citenamefont {Watkins}, \citenamefont
  {Freeman}, \citenamefont {Murphy}, \citenamefont {Afanasyev}, \citenamefont
  {Buldyrev}, \citenamefont {Luz}, \citenamefont {Raposo}, \citenamefont
  {Stanley},\ and\ \citenamefont {Viswanathan}}]{Edwards:2007p598}%
  \BibitemOpen
  \bibfield  {author} {\bibinfo {author} {\bibfnamefont {A.~M.}\ \bibnamefont
  {Edwards}}, \bibinfo {author} {\bibfnamefont {R.~A.}\ \bibnamefont
  {Phillips}}, \bibinfo {author} {\bibfnamefont {N.~W.}\ \bibnamefont
  {Watkins}}, \bibinfo {author} {\bibfnamefont {M.~P.}\ \bibnamefont
  {Freeman}}, \bibinfo {author} {\bibfnamefont {E.~J.}\ \bibnamefont {Murphy}},
  \bibinfo {author} {\bibfnamefont {V.}~\bibnamefont {Afanasyev}}, \bibinfo
  {author} {\bibfnamefont {S.~V.}\ \bibnamefont {Buldyrev}}, \bibinfo {author}
  {\bibfnamefont {M.~G. E.~D.}\ \bibnamefont {Luz}}, \bibinfo {author}
  {\bibfnamefont {E.~P.}\ \bibnamefont {Raposo}}, \bibinfo {author}
  {\bibfnamefont {H.~E.}\ \bibnamefont {Stanley}}, \ and\ \bibinfo {author}
  {\bibfnamefont {G.~M.}\ \bibnamefont {Viswanathan}},\ }\href {\doibase
  10.1038/nature06199} {\bibfield  {journal} {\bibinfo  {journal} {Nature}\
  }\textbf {\bibinfo {volume} {449}},\ \bibinfo {pages} {1044} (\bibinfo {year}
  {2007})}\BibitemShut {NoStop}%
\bibitem [{\citenamefont {Helbing}(2001)}]{RevModPhys.73.1067}%
  \BibitemOpen
  \bibfield  {author} {\bibinfo {author} {\bibfnamefont {D.}~\bibnamefont
  {Helbing}},\ }\href@noop {} {\bibfield  {journal} {\bibinfo  {journal} {Rev.
  Mod. Phys.}\ }\textbf {\bibinfo {volume} {73}},\ \bibinfo {pages} {1067}
  (\bibinfo {year} {2001})}\BibitemShut {NoStop}%
\bibitem [{\citenamefont {Aw}\ \emph {et~al.}(2002)\citenamefont {Aw},
  \citenamefont {Klar}, \citenamefont {Materne},\ and\ \citenamefont
  {Rascle}}]{Aw2002SIAM259}%
  \BibitemOpen
  \bibfield  {author} {\bibinfo {author} {\bibfnamefont {A.}~\bibnamefont
  {Aw}}, \bibinfo {author} {\bibfnamefont {A.}~\bibnamefont {Klar}}, \bibinfo
  {author} {\bibfnamefont {T.}~\bibnamefont {Materne}}, \ and\ \bibinfo
  {author} {\bibfnamefont {M.}~\bibnamefont {Rascle}},\ }\href@noop {}
  {\bibfield  {journal} {\bibinfo  {journal} {SIAM J. Appl. Math.}\ }\textbf
  {\bibinfo {volume} {63}},\ \bibinfo {pages} {259} (\bibinfo {year}
  {2002})}\BibitemShut {NoStop}%
\bibitem [{\citenamefont {Nourhani}\ \emph {et~al.}(2014)\citenamefont
  {Nourhani}, \citenamefont {Lammert}, \citenamefont {Borhan},\ and\
  \citenamefont {Crespi}}]{Nourhani2014PRE062304}%
  \BibitemOpen
  \bibfield  {author} {\bibinfo {author} {\bibfnamefont {A.}~\bibnamefont
  {Nourhani}}, \bibinfo {author} {\bibfnamefont {P.~E.}\ \bibnamefont
  {Lammert}}, \bibinfo {author} {\bibfnamefont {A.}~\bibnamefont {Borhan}}, \
  and\ \bibinfo {author} {\bibfnamefont {V.~H.}\ \bibnamefont {Crespi}},\
  }\href@noop {} {\bibfield  {journal} {\bibinfo  {journal} {Phys. Rev. E}\
  }\textbf {\bibinfo {volume} {89}},\ \bibinfo {pages} {062304} (\bibinfo
  {year} {2014})}\BibitemShut {NoStop}%
\bibitem [{\citenamefont {Ebbens}\ \emph {et~al.}(2010)\citenamefont {Ebbens},
  \citenamefont {Jones}, \citenamefont {Ryan}, \citenamefont {Golestanian},\
  and\ \citenamefont {Howse}}]{Ebbens:2010p589}%
  \BibitemOpen
  \bibfield  {author} {\bibinfo {author} {\bibfnamefont {S.}~\bibnamefont
  {Ebbens}}, \bibinfo {author} {\bibfnamefont {R.~A.~L.}\ \bibnamefont
  {Jones}}, \bibinfo {author} {\bibfnamefont {A.~J.}\ \bibnamefont {Ryan}},
  \bibinfo {author} {\bibfnamefont {R.}~\bibnamefont {Golestanian}}, \ and\
  \bibinfo {author} {\bibfnamefont {J.~R.}\ \bibnamefont {Howse}},\ }\href
  {\doibase 10.1103/PhysRevE.82.015304} {\bibfield  {journal} {\bibinfo
  {journal} {Phys. Rev. E}\ }\textbf {\bibinfo {volume} {82}},\ \bibinfo
  {pages} {015304} (\bibinfo {year} {2010})}\BibitemShut {NoStop}%
\bibitem [{\citenamefont {Takagi}\ \emph {et~al.}(2013)\citenamefont {Takagi},
  \citenamefont {Braunschweig}, \citenamefont {Zhang},\ and\ \citenamefont
  {Shelley}}]{Takagi:2013p645}%
  \BibitemOpen
  \bibfield  {author} {\bibinfo {author} {\bibfnamefont {D.}~\bibnamefont
  {Takagi}}, \bibinfo {author} {\bibfnamefont {A.~B.}\ \bibnamefont
  {Braunschweig}}, \bibinfo {author} {\bibfnamefont {J.}~\bibnamefont {Zhang}},
  \ and\ \bibinfo {author} {\bibfnamefont {M.~J.}\ \bibnamefont {Shelley}},\
  }\href {\doibase 10.1103/PhysRevLett.110.038301} {\bibfield  {journal}
  {\bibinfo  {journal} {Phys. Rev. Lett.}\ }\textbf {\bibinfo {volume} {110}},\
  \bibinfo {pages} {038301} (\bibinfo {year} {2013})}\BibitemShut {NoStop}%
\bibitem [{\citenamefont {Nourhani}\ \emph
  {et~al.}(2013{\natexlab{b}})\citenamefont {Nourhani}, \citenamefont
  {Lammert}, \citenamefont {Borhan},\ and\ \citenamefont
  {Crespi}}]{Nourhani2013p050301}%
  \BibitemOpen
  \bibfield  {author} {\bibinfo {author} {\bibfnamefont {A.}~\bibnamefont
  {Nourhani}}, \bibinfo {author} {\bibfnamefont {P.~E.}\ \bibnamefont
  {Lammert}}, \bibinfo {author} {\bibfnamefont {A.}~\bibnamefont {Borhan}}, \
  and\ \bibinfo {author} {\bibfnamefont {V.~H.}\ \bibnamefont {Crespi}},\
  }\href@noop {} {\bibfield  {journal} {\bibinfo  {journal} {Phys. Rev. E}\
  }\textbf {\bibinfo {volume} {87}},\ \bibinfo {pages} {050301(R)} (\bibinfo
  {year} {2013}{\natexlab{b}})}\BibitemShut {NoStop}%
\bibitem [{\citenamefont {Teeffelen}\ and\ \citenamefont
  {L{\"o}wen}(2008)}]{VanTeeffelen:2008p643}%
  \BibitemOpen
  \bibfield  {author} {\bibinfo {author} {\bibfnamefont {S.~V.}\ \bibnamefont
  {Teeffelen}}\ and\ \bibinfo {author} {\bibfnamefont {H.}~\bibnamefont
  {L{\"o}wen}},\ }\href {\doibase 10.1103/PhysRevE.78.020101} {\bibfield
  {journal} {\bibinfo  {journal} {Phys. Rev. E}\ }\textbf {\bibinfo {volume}
  {78}},\ \bibinfo {pages} {020101} (\bibinfo {year} {2008})}\BibitemShut
  {NoStop}%
\bibitem [{\citenamefont {Schweitzer}\ \emph {et~al.}(1998)\citenamefont
  {Schweitzer}, \citenamefont {Ebeling},\ and\ \citenamefont
  {Tilch}}]{Schweitzer1998PRL5044}%
  \BibitemOpen
  \bibfield  {author} {\bibinfo {author} {\bibfnamefont {F.}~\bibnamefont
  {Schweitzer}}, \bibinfo {author} {\bibfnamefont {W.}~\bibnamefont {Ebeling}},
  \ and\ \bibinfo {author} {\bibfnamefont {B.}~\bibnamefont {Tilch}},\
  }\href@noop {} {\bibfield  {journal} {\bibinfo  {journal} {Phys. Rev. Lett.}\
  }\textbf {\bibinfo {volume} {80}},\ \bibinfo {pages} {5044} (\bibinfo {year}
  {1998})}\BibitemShut {NoStop}%
\bibitem [{\citenamefont {Cates}\ and\ \citenamefont
  {Tailleur}(2013)}]{Cates:2013p620}%
  \BibitemOpen
  \bibfield  {author} {\bibinfo {author} {\bibfnamefont {M.~E.}\ \bibnamefont
  {Cates}}\ and\ \bibinfo {author} {\bibfnamefont {J.}~\bibnamefont
  {Tailleur}},\ }\href {\doibase 10.1209/0295-5075/101/20010} {\bibfield
  {journal} {\bibinfo  {journal} {EPL}\ }\textbf {\bibinfo {volume} {101}},\
  \bibinfo {pages} {20010} (\bibinfo {year} {2013})}\BibitemShut {NoStop}%
\bibitem [{\citenamefont {Romanczuk}\ \emph {et~al.}(2009)\citenamefont
  {Romanczuk}, \citenamefont {Couzin},\ and\ \citenamefont
  {Schimansky-Geier}}]{Romanczuk:2009p659}%
  \BibitemOpen
  \bibfield  {author} {\bibinfo {author} {\bibfnamefont {P.}~\bibnamefont
  {Romanczuk}}, \bibinfo {author} {\bibfnamefont {I.}~\bibnamefont {Couzin}}, \
  and\ \bibinfo {author} {\bibfnamefont {L.}~\bibnamefont {Schimansky-Geier}},\
  }\href {\doibase 10.1103/PhysRevLett.102.010602} {\bibfield  {journal}
  {\bibinfo  {journal} {Phys. Rev. Lett.}\ }\textbf {\bibinfo {volume} {102}},\
  \bibinfo {pages} {010602} (\bibinfo {year} {2009})}\BibitemShut {NoStop}%
\bibitem [{\citenamefont {Mandal}\ and\ \citenamefont
  {Ghosh}(2013)}]{Mandal2013PRL248101}%
  \BibitemOpen
  \bibfield  {author} {\bibinfo {author} {\bibfnamefont {P.}~\bibnamefont
  {Mandal}}\ and\ \bibinfo {author} {\bibfnamefont {A.}~\bibnamefont {Ghosh}},\
  }\href@noop {} {\bibfield  {journal} {\bibinfo  {journal} {Phys. Rev. Lett.}\
  }\textbf {\bibinfo {volume} {111}},\ \bibinfo {pages} {248101} (\bibinfo
  {year} {2013})}\BibitemShut {NoStop}%
\bibitem [{\citenamefont {Hanggi}\ and\ \citenamefont
  {Jung}(1995)}]{Hanggi1995ACP239}%
  \BibitemOpen
  \bibfield  {author} {\bibinfo {author} {\bibfnamefont {P.}~\bibnamefont
  {Hanggi}}\ and\ \bibinfo {author} {\bibfnamefont {P.}~\bibnamefont {Jung}},\
  }\href@noop {} {\bibfield  {journal} {\bibinfo  {journal} {Adv. Che. Phys.}\
  }\textbf {\bibinfo {volume} {LXXXIX}},\ \bibinfo {pages} {239} (\bibinfo
  {year} {1995})}\BibitemShut {NoStop}%
\bibitem [{\citenamefont {K{\l}osek-Dygas}\ \emph {et~al.}(1988)\citenamefont
  {K{\l}osek-Dygas}, \citenamefont {Matkowsky},\ and\ \citenamefont
  {Schuss}}]{KlosekDygas1988SIAM425}%
  \BibitemOpen
  \bibfield  {author} {\bibinfo {author} {\bibfnamefont {M.~M.}\ \bibnamefont
  {K{\l}osek-Dygas}}, \bibinfo {author} {\bibfnamefont {B.~J.}\ \bibnamefont
  {Matkowsky}}, \ and\ \bibinfo {author} {\bibfnamefont {Z.}~\bibnamefont
  {Schuss}},\ }\href@noop {} {\bibfield  {journal} {\bibinfo  {journal} {SIAM
  Journal on Applied Mathematics}\ }\textbf {\bibinfo {volume} {48}},\ \bibinfo
  {pages} {425} (\bibinfo {year} {1988})}\BibitemShut {NoStop}%
\bibitem [{\citenamefont {San~Miguel}\ and\ \citenamefont
  {Sancho}(1980)}]{SanMiguel1980JSP605}%
  \BibitemOpen
  \bibfield  {author} {\bibinfo {author} {\bibfnamefont {M.}~\bibnamefont
  {San~Miguel}}\ and\ \bibinfo {author} {\bibfnamefont {J.~M.}\ \bibnamefont
  {Sancho}},\ }\href@noop {} {\bibfield  {journal} {\bibinfo  {journal} {J.
  Stat. Phys.}\ }\textbf {\bibinfo {volume} {22}},\ \bibinfo {pages} {605}
  (\bibinfo {year} {1980})}\BibitemShut {NoStop}%
\bibitem [{\citenamefont {Kamenev}\ \emph {et~al.}(2008)\citenamefont
  {Kamenev}, \citenamefont {Meerson},\ and\ \citenamefont
  {Shklovskii}}]{Kamenev2008PRL268103}%
  \BibitemOpen
  \bibfield  {author} {\bibinfo {author} {\bibfnamefont {A.}~\bibnamefont
  {Kamenev}}, \bibinfo {author} {\bibfnamefont {B.}~\bibnamefont {Meerson}}, \
  and\ \bibinfo {author} {\bibfnamefont {B.}~\bibnamefont {Shklovskii}},\
  }\href@noop {} {\bibfield  {journal} {\bibinfo  {journal} {Phys. Rev. Lett.}\
  }\textbf {\bibinfo {volume} {101}},\ \bibinfo {pages} {268103} (\bibinfo
  {year} {2008})}\BibitemShut {NoStop}%
\bibitem [{\citenamefont {Peruani}\ and\ \citenamefont
  {Morelli}(2007)}]{Peruani:2007p602}%
  \BibitemOpen
  \bibfield  {author} {\bibinfo {author} {\bibfnamefont {F.}~\bibnamefont
  {Peruani}}\ and\ \bibinfo {author} {\bibfnamefont {L.}~\bibnamefont
  {Morelli}},\ }\href {\doibase 10.1103/PhysRevLett.99.010602} {\bibfield
  {journal} {\bibinfo  {journal} {Phys. Rev. Lett.}\ }\textbf {\bibinfo
  {volume} {99}},\ \bibinfo {pages} {010602} (\bibinfo {year}
  {2007})}\BibitemShut {NoStop}%
\bibitem [{\citenamefont {Lauga}\ \emph {et~al.}(2006)\citenamefont {Lauga},
  \citenamefont {Diluzio}, \citenamefont {Whitesides},\ and\ \citenamefont
  {Stone}}]{Lauga:2006p606}%
  \BibitemOpen
  \bibfield  {author} {\bibinfo {author} {\bibfnamefont {E.}~\bibnamefont
  {Lauga}}, \bibinfo {author} {\bibfnamefont {W.~R.}\ \bibnamefont {Diluzio}},
  \bibinfo {author} {\bibfnamefont {G.~M.}\ \bibnamefont {Whitesides}}, \ and\
  \bibinfo {author} {\bibfnamefont {H.~A.}\ \bibnamefont {Stone}},\ }\href
  {\doibase 10.1529/biophysj.105.069401} {\bibfield  {journal} {\bibinfo
  {journal} {Biophys. J.}\ }\textbf {\bibinfo {volume} {90}},\ \bibinfo {pages}
  {400} (\bibinfo {year} {2006})}\BibitemShut {NoStop}%
\bibitem [{\citenamefont {Steinberger}\ \emph {et~al.}(1994)\citenamefont
  {Steinberger}, \citenamefont {Petersen}, \citenamefont {Petermann},\ and\
  \citenamefont {Weiss}}]{FLM:378272}%
  \BibitemOpen
  \bibfield  {author} {\bibinfo {author} {\bibfnamefont {B.}~\bibnamefont
  {Steinberger}}, \bibinfo {author} {\bibfnamefont {N.}~\bibnamefont
  {Petersen}}, \bibinfo {author} {\bibfnamefont {H.}~\bibnamefont {Petermann}},
  \ and\ \bibinfo {author} {\bibfnamefont {D.~G.}\ \bibnamefont {Weiss}},\
  }\href {\doibase 10.1017/S0022112094001904} {\bibfield  {journal} {\bibinfo
  {journal} {Journal of Fluid Mechanics}\ }\textbf {\bibinfo {volume} {273}},\
  \bibinfo {pages} {189} (\bibinfo {year} {1994})}\BibitemShut {NoStop}%
\bibitem [{\citenamefont {C\ifmmode~\bar{e}\else \={e}\fi{}bers}\ and\
  \citenamefont {Ozols}(2006)}]{PhysRevE.73.021505}%
  \BibitemOpen
  \bibfield  {author} {\bibinfo {author} {\bibfnamefont {A.}~\bibnamefont
  {C\ifmmode~\bar{e}\else \={e}\fi{}bers}}\ and\ \bibinfo {author}
  {\bibfnamefont {M.}~\bibnamefont {Ozols}},\ }\href {\doibase
  10.1103/PhysRevE.73.021505} {\bibfield  {journal} {\bibinfo  {journal} {Phys.
  Rev. E}\ }\textbf {\bibinfo {volume} {73}},\ \bibinfo {pages} {021505}
  (\bibinfo {year} {2006})}\BibitemShut {NoStop}%
\bibitem [{\citenamefont {{\=E}rglis}\ \emph {et~al.}(2007)\citenamefont
  {{\=E}rglis}, \citenamefont {Wen}, \citenamefont {Ose}, \citenamefont
  {Zeltins}, \citenamefont {Sharipo}, \citenamefont {Janmey},\ and\
  \citenamefont {C{\=e}bers}}]{Erglis:2007p596}%
  \BibitemOpen
  \bibfield  {author} {\bibinfo {author} {\bibfnamefont {K.}~\bibnamefont
  {{\=E}rglis}}, \bibinfo {author} {\bibfnamefont {Q.}~\bibnamefont {Wen}},
  \bibinfo {author} {\bibfnamefont {V.}~\bibnamefont {Ose}}, \bibinfo {author}
  {\bibfnamefont {A.}~\bibnamefont {Zeltins}}, \bibinfo {author} {\bibfnamefont
  {A.}~\bibnamefont {Sharipo}}, \bibinfo {author} {\bibfnamefont {P.~A.}\
  \bibnamefont {Janmey}}, \ and\ \bibinfo {author} {\bibfnamefont
  {A.}~\bibnamefont {C{\=e}bers}},\ }\href {\doibase
  10.1529/biophysj.107.107474} {\bibfield  {journal} {\bibinfo  {journal}
  {Biophys. J.}\ }\textbf {\bibinfo {volume} {93}},\ \bibinfo {pages} {1402}
  (\bibinfo {year} {2007})}\BibitemShut {NoStop}%
\bibitem [{\citenamefont {C{\=e}bers}(2011)}]{Cebers:2011p639}%
  \BibitemOpen
  \bibfield  {author} {\bibinfo {author} {\bibfnamefont {A.}~\bibnamefont
  {C{\=e}bers}},\ }\href {\doibase 10.1016/j.jmmm.2010.09.017} {\bibfield
  {journal} {\bibinfo  {journal} {Journal of Magnetism and Magnetic Materials}\
  }\textbf {\bibinfo {volume} {323}},\ \bibinfo {pages} {279} (\bibinfo {year}
  {2011})}\BibitemShut {NoStop}%
\bibitem [{\citenamefont {Brydges}\ \emph {et~al.}(1982)\citenamefont
  {Brydges}, \citenamefont {Frohlich},\ and\ \citenamefont
  {Spencer}}]{Brydges1982CMP123}%
  \BibitemOpen
  \bibfield  {author} {\bibinfo {author} {\bibfnamefont {D.}~\bibnamefont
  {Brydges}}, \bibinfo {author} {\bibfnamefont {J.}~\bibnamefont {Frohlich}}, \
  and\ \bibinfo {author} {\bibfnamefont {T.}~\bibnamefont {Spencer}},\
  }\href@noop {} {\bibfield  {journal} {\bibinfo  {journal} {Commun. Math.
  Phys.}\ }\textbf {\bibinfo {volume} {83}},\ \bibinfo {pages} {123} (\bibinfo
  {year} {1982})}\BibitemShut {NoStop}%
\bibitem [{\citenamefont {Strefler}\ \emph {et~al.}(2009)\citenamefont
  {Strefler}, \citenamefont {Ebeling}, \citenamefont {Gudowska-Nowak},\ and\
  \citenamefont {Schimansky-Geier}}]{Strefler:2009p605}%
  \BibitemOpen
  \bibfield  {author} {\bibinfo {author} {\bibfnamefont {J.}~\bibnamefont
  {Strefler}}, \bibinfo {author} {\bibfnamefont {W.}~\bibnamefont {Ebeling}},
  \bibinfo {author} {\bibfnamefont {E.}~\bibnamefont {Gudowska-Nowak}}, \ and\
  \bibinfo {author} {\bibfnamefont {L.}~\bibnamefont {Schimansky-Geier}},\
  }\href {\doibase 10.1140/epjb/e2009-00408-8} {\bibfield  {journal} {\bibinfo
  {journal} {Eur. Phys. J. B}\ }\textbf {\bibinfo {volume} {72}},\ \bibinfo
  {pages} {597} (\bibinfo {year} {2009})}\BibitemShut {NoStop}%
\bibitem [{\citenamefont {Reichhardt}\ and\ \citenamefont
  {Reichhardt}(2013)}]{Reichhardt2013PRE042306}%
  \BibitemOpen
  \bibfield  {author} {\bibinfo {author} {\bibfnamefont {C.}~\bibnamefont
  {Reichhardt}}\ and\ \bibinfo {author} {\bibfnamefont {C.~J.~O.}\ \bibnamefont
  {Reichhardt}},\ }\href@noop {} {\bibfield  {journal} {\bibinfo  {journal}
  {Phys. Rev. E}\ }\textbf {\bibinfo {volume} {88}},\ \bibinfo {pages} {042306}
  (\bibinfo {year} {2013})}\BibitemShut {NoStop}%
\bibitem [{\citenamefont {Ornstein}\ and\ \citenamefont
  {Uhlenbeck}(1930)}]{OUoriginal1930}%
  \BibitemOpen
  \bibfield  {author} {\bibinfo {author} {\bibfnamefont {G.~E.}\ \bibnamefont
  {Ornstein}}\ and\ \bibinfo {author} {\bibfnamefont {L.~S.}\ \bibnamefont
  {Uhlenbeck}},\ }\href@noop {} {\bibfield  {journal} {\bibinfo  {journal}
  {Phys. Rev.}\ }\textbf {\bibinfo {volume} {36}},\ \bibinfo {pages} {679}
  (\bibinfo {year} {1930})}\BibitemShut {NoStop}%
\bibitem [{\citenamefont {Weber}\ \emph {et~al.}(2011)\citenamefont {Weber},
  \citenamefont {Radtke}, \citenamefont {Schimansky-Geier},\ and\ \citenamefont
  {H{\"a}nggi}}]{Weber:2011p644}%
  \BibitemOpen
  \bibfield  {author} {\bibinfo {author} {\bibfnamefont {C.}~\bibnamefont
  {Weber}}, \bibinfo {author} {\bibfnamefont {P.~K.}\ \bibnamefont {Radtke}},
  \bibinfo {author} {\bibfnamefont {L.}~\bibnamefont {Schimansky-Geier}}, \
  and\ \bibinfo {author} {\bibfnamefont {P.}~\bibnamefont {H{\"a}nggi}},\
  }\href {\doibase 10.1103/PhysRevE.84.011132} {\bibfield  {journal} {\bibinfo
  {journal} {Phys. Rev. E}\ }\textbf {\bibinfo {volume} {84}},\ \bibinfo
  {pages} {011132} (\bibinfo {year} {2011})}\BibitemShut {NoStop}%
\bibitem [{\citenamefont {Dieterich}\ \emph {et~al.}(2008)\citenamefont
  {Dieterich}, \citenamefont {Klages}, \citenamefont {Preuss},\ and\
  \citenamefont {Schwab}}]{Dieterich2008PNAS459}%
  \BibitemOpen
  \bibfield  {author} {\bibinfo {author} {\bibfnamefont {P.}~\bibnamefont
  {Dieterich}}, \bibinfo {author} {\bibfnamefont {R.}~\bibnamefont {Klages}},
  \bibinfo {author} {\bibfnamefont {R.}~\bibnamefont {Preuss}}, \ and\ \bibinfo
  {author} {\bibfnamefont {A.}~\bibnamefont {Schwab}},\ }\href@noop {}
  {\bibfield  {journal} {\bibinfo  {journal} {Proc. Natl. Acad. Sci. USA}\
  }\textbf {\bibinfo {volume} {105}},\ \bibinfo {pages} {459} (\bibinfo {year}
  {2008})}\BibitemShut {NoStop}%
\bibitem [{\citenamefont {Radtke}\ and\ \citenamefont
  {Schimansky-Geier}(2012)}]{Radtke2012PRE051110}%
  \BibitemOpen
  \bibfield  {author} {\bibinfo {author} {\bibfnamefont {P.~K.}\ \bibnamefont
  {Radtke}}\ and\ \bibinfo {author} {\bibfnamefont {L.}~\bibnamefont
  {Schimansky-Geier}},\ }\href@noop {} {\bibfield  {journal} {\bibinfo
  {journal} {Phys. Rev. E}\ }\textbf {\bibinfo {volume} {85}},\ \bibinfo
  {pages} {051110} (\bibinfo {year} {2012})}\BibitemShut {NoStop}%
\bibitem [{\citenamefont {Yushchenko}\ and\ \citenamefont
  {Yu.~Badalyan}(2012)}]{Yushchenko2012PRE051127}%
  \BibitemOpen
  \bibfield  {author} {\bibinfo {author} {\bibfnamefont {O.~V.}\ \bibnamefont
  {Yushchenko}}\ and\ \bibinfo {author} {\bibfnamefont {A.}~\bibnamefont
  {Yu.~Badalyan}},\ }\href@noop {} {\bibfield  {journal} {\bibinfo  {journal}
  {Phys. Rev. E}\ }\textbf {\bibinfo {volume} {85}},\ \bibinfo {pages} {051127}
  (\bibinfo {year} {2012})}\BibitemShut {NoStop}%
\bibitem [{\citenamefont {Szamel}(2014)}]{Szamel2014PRE012111}%
  \BibitemOpen
  \bibfield  {author} {\bibinfo {author} {\bibfnamefont {G.}~\bibnamefont
  {Szamel}},\ }\href@noop {} {\bibfield  {journal} {\bibinfo  {journal} {Phys.
  Rev. E}\ }\textbf {\bibinfo {volume} {90}},\ \bibinfo {pages} {012111}
  (\bibinfo {year} {2014})}\BibitemShut {NoStop}%
\bibitem [{\citenamefont {Torney}\ and\ \citenamefont
  {Neufeld}(2008)}]{Torney2008PRL078105}%
  \BibitemOpen
  \bibfield  {author} {\bibinfo {author} {\bibfnamefont {C.}~\bibnamefont
  {Torney}}\ and\ \bibinfo {author} {\bibfnamefont {Z.}~\bibnamefont
  {Neufeld}},\ }\href@noop {} {\bibfield  {journal} {\bibinfo  {journal} {Phys.
  Rev. Lett.}\ }\textbf {\bibinfo {volume} {101}},\ \bibinfo {pages} {078105}
  (\bibinfo {year} {2008})}\BibitemShut {NoStop}%
\bibitem [{\citenamefont {Olsen}(2013)}]{Olsen2013PRA051802}%
  \BibitemOpen
  \bibfield  {author} {\bibinfo {author} {\bibfnamefont {M.~K.}\ \bibnamefont
  {Olsen}},\ }\href@noop {} {\bibfield  {journal} {\bibinfo  {journal} {Phys.
  Rev. A}\ }\textbf {\bibinfo {volume} {88}},\ \bibinfo {pages} {051802(R)}
  (\bibinfo {year} {2013})}\BibitemShut {NoStop}%
\bibitem [{\citenamefont {Sarovar}\ and\ \citenamefont
  {Grace}(2012)}]{Sarovar2012PRL130401}%
  \BibitemOpen
  \bibfield  {author} {\bibinfo {author} {\bibfnamefont {M.}~\bibnamefont
  {Sarovar}}\ and\ \bibinfo {author} {\bibfnamefont {M.~D.}\ \bibnamefont
  {Grace}},\ }\href@noop {} {\bibfield  {journal} {\bibinfo  {journal} {Phys.
  Rev. Lett.}\ }\textbf {\bibinfo {volume} {109}},\ \bibinfo {pages} {130401}
  (\bibinfo {year} {2012})}\BibitemShut {NoStop}%
\bibitem [{\citenamefont {Jing}\ and\ \citenamefont
  {Yu}(2010)}]{Jing2010PRL40403}%
  \BibitemOpen
  \bibfield  {author} {\bibinfo {author} {\bibfnamefont {J.}~\bibnamefont
  {Jing}}\ and\ \bibinfo {author} {\bibfnamefont {T.}~\bibnamefont {Yu}},\
  }\href@noop {} {\bibfield  {journal} {\bibinfo  {journal} {Phys. Rev. Lett.}\
  }\textbf {\bibinfo {volume} {105}},\ \bibinfo {pages} {240403} (\bibinfo
  {year} {2010})}\BibitemShut {NoStop}%
\bibitem [{\citenamefont {Stimming}\ \emph {et~al.}(2010)\citenamefont
  {Stimming}, \citenamefont {Mauser}, \citenamefont {Schmiedmayer},\ and\
  \citenamefont {Mazets}}]{Stimming02010PRL15301}%
  \BibitemOpen
  \bibfield  {author} {\bibinfo {author} {\bibfnamefont {H.-P.}\ \bibnamefont
  {Stimming}}, \bibinfo {author} {\bibfnamefont {N.~J.}\ \bibnamefont
  {Mauser}}, \bibinfo {author} {\bibfnamefont {J.}~\bibnamefont
  {Schmiedmayer}}, \ and\ \bibinfo {author} {\bibfnamefont {I.~E.}\
  \bibnamefont {Mazets}},\ }\href@noop {} {\bibfield  {journal} {\bibinfo
  {journal} {Phys. Rev. Lett.}\ }\textbf {\bibinfo {volume} {105}},\ \bibinfo
  {pages} {015301} (\bibinfo {year} {2010})}\BibitemShut {NoStop}%
\bibitem [{\citenamefont {Hu}\ \emph {et~al.}(2014)\citenamefont {Hu},
  \citenamefont {Trousdale}, \citenamefont {Josic},\ and\ \citenamefont
  {Shea-Brown}}]{Hu2014PRE032802}%
  \BibitemOpen
  \bibfield  {author} {\bibinfo {author} {\bibfnamefont {Y.}~\bibnamefont
  {Hu}}, \bibinfo {author} {\bibfnamefont {J.}~\bibnamefont {Trousdale}},
  \bibinfo {author} {\bibfnamefont {K.}~\bibnamefont {Josic}}, \ and\ \bibinfo
  {author} {\bibfnamefont {E.}~\bibnamefont {Shea-Brown}},\ }\href@noop {}
  {\bibfield  {journal} {\bibinfo  {journal} {Phys. Rev. E}\ }\textbf {\bibinfo
  {volume} {89}},\ \bibinfo {pages} {032802} (\bibinfo {year}
  {2014})}\BibitemShut {NoStop}%
\bibitem [{\citenamefont {Assaf}\ \emph {et~al.}(2013)\citenamefont {Assaf},
  \citenamefont {Roberts}, \citenamefont {Luthey-Schulten},\ and\ \citenamefont
  {Goldenfeld}}]{Assaf2013PRL058102}%
  \BibitemOpen
  \bibfield  {author} {\bibinfo {author} {\bibfnamefont {M.}~\bibnamefont
  {Assaf}}, \bibinfo {author} {\bibfnamefont {E.}~\bibnamefont {Roberts}},
  \bibinfo {author} {\bibfnamefont {Z.}~\bibnamefont {Luthey-Schulten}}, \ and\
  \bibinfo {author} {\bibfnamefont {N.}~\bibnamefont {Goldenfeld}},\
  }\href@noop {} {\bibfield  {journal} {\bibinfo  {journal} {Phys. Rev. Lett.}\
  }\textbf {\bibinfo {volume} {111}},\ \bibinfo {pages} {058102} (\bibinfo
  {year} {2013})}\BibitemShut {NoStop}%
\bibitem [{\citenamefont {Charlebois}\ \emph {et~al.}(2011)\citenamefont
  {Charlebois}, \citenamefont {Abdennur},\ and\ \citenamefont
  {Kaern}}]{Charlebois2011PRL218101}%
  \BibitemOpen
  \bibfield  {author} {\bibinfo {author} {\bibfnamefont {D.~A.}\ \bibnamefont
  {Charlebois}}, \bibinfo {author} {\bibfnamefont {N.}~\bibnamefont
  {Abdennur}}, \ and\ \bibinfo {author} {\bibfnamefont {M.}~\bibnamefont
  {Kaern}},\ }\href@noop {} {\bibfield  {journal} {\bibinfo  {journal} {Phys.
  Rev. Lett.}\ }\textbf {\bibinfo {volume} {107}},\ \bibinfo {pages} {218101}
  (\bibinfo {year} {2011})}\BibitemShut {NoStop}%
\bibitem [{\citenamefont {Berg}(2008)}]{Berg2008PRL188101}%
  \BibitemOpen
  \bibfield  {author} {\bibinfo {author} {\bibfnamefont {J.}~\bibnamefont
  {Berg}},\ }\href@noop {} {\bibfield  {journal} {\bibinfo  {journal} {Phys.
  Rev. Lett.}\ }\textbf {\bibinfo {volume} {100}},\ \bibinfo {pages} {188101}
  (\bibinfo {year} {2008})}\BibitemShut {NoStop}%
\end{thebibliography}
\end{document}